\documentclass[journal=jctcce,manuscript=article]{achemso}
\setkeys{acs}{biblabel=brackets}
\setkeys{acs}{usetitle=true}
\setkeys{acs}{maxauthors = 0}

\usepackage{amsmath}
\usepackage{caption}
\usepackage{subcaption}
\usepackage[hidelinks]{hyperref}
\usepackage{tikz}
\usetikzlibrary{positioning,shapes,arrows,backgrounds,calc,fit}

\newcommand{\onlinecite}[1]{\hspace{-1 ex} \nocite{#1}\citenum{#1}}
\newcommand{\br}{\mathbf{r}}
\newcommand{\bs}{\mathbf{s}}
\newcommand{\bS}{\mathbf{S}}
\newcommand{\bt}{\mathbf{t}}
\newcommand{\bV}{\mathbf{V}}
\newcommand{\bra}[1]{\langle #1|}
\newcommand{\braket}[2]{\langle #1|#2\rangle}
\newcommand{\ket}[1]{|#1\rangle}
\newcommand{\nline}{\nonumber \\}
\newcommand{\Trace}[1]{\mathrm{Tr}[#1]}

\newcommand{\edit}[1]{#1}

\definecolor{seabornbggrey}{HTML}{d2d2d9} 
\definecolor{seabornblue}{HTML}{4c72b0}
\definecolor{seabornred}{HTML}{c44e52}
\tikzstyle{box0} = [rectangle, fill=seabornred, text centered, text=white, font=\footnotesize, font=\sffamily]
\tikzstyle{box1} = [rectangle, fill=seabornblue, text centered, text=white,   font=\footnotesize, font=\sffamily]
\tikzstyle{box2} = [rectangle, fill=seabornbggrey, text centered, text=black,   font=\footnotesize, font=\sffamily]
\tikzstyle{boxwhite} = [rectangle, fill=white, text centered, text=black,   font=\footnotesize, font=\sffamily, rounded corners=0.1cm]
\tikzstyle{dummy} = [font=\sffamily]
\tikzstyle{line} = [draw, ultra thick, -latex']

\author{Edward B. Linscott}
\affiliation[École polytechnique fédérale de Lausanne]
{Theory and Simulation of Materials (THEOS), \'Ecole Polytechnique F\'ed\'erale de Lausanne, 1015 Lausanne, Switzerland}
\author{Daniel J. Cole}
\affiliation[Newcastle University]
{School of Natural and Environmental Sciences, Newcastle University, Newcastle upon Tyne NE1 7RU, United Kingdom}
\author{Nicholas D. M. Hine}
\affiliation[University of Warwick]{Department of Physics, University of Warwick, Coventry CV4 7AL, United Kingdom}
\author{Michael C. Payne}
\affiliation[University of Cambridge]
{Theory of Condensed Matter, Cavendish Laboratory, University of Cambridge, 19 JJ Thomson Ave, Cambridge CB3 0HE, United Kingdom}
\author{C\'edric Weber}
\affiliation[King's College London]
{Theory and Simulation of Condensed Matter, King's College London, The Strand, London WC2R 2LS, United Kingdom}
\email{cedric.weber@kcl.ac.uk}

\title[]
{ONETEP + TOSCAM: uniting dynamical mean field theory and linear-scaling density functional theory}

\begin{document}

\clearpage

\begin{abstract}
We introduce the unification of dynamical mean field theory (DMFT) and linear-scaling density functional theory (DFT), as recently implemented in ONETEP, a linear-scaling DFT package, and TOSCAM, a DMFT toolbox. This code can account for strongly correlated electronic behavior while simultaneously including the effects of the environment, making it ideally suited for studying complex and heterogeneous systems that contain transition metals and lanthanides, such as metalloproteins. We systematically introduce the necessary formalism, which must account for the non-orthogonal basis set used by ONETEP. In order to demonstrate the capabilities of this code, we apply it to carbon monoxide-ligated iron porphyrin and explore the distinctly quantum-mechanical character of the iron $3d$ electrons during the process of photodissociation.

\end{abstract}

\clearpage

\section{Introduction}

In the last few decades, density functional theory (DFT) has established itself as a key method in computational materials science \cite{Hohenberg1964a, Kohn1965a, Jones2015a, Jain2016a}. Facilitated by exponentially increasing computing power, modern DFT codes are capable of routinely calculating the electronic structure of hundreds of atoms, opening the door to quantum-mechanical modeling of a vast landscape of systems of considerable scientific interest.

The range of computationally accessible systems has broadened even further with the advent of linear-scaling DFT codes (that is, codes whose computational cost scales linearly with the number of atoms in the system, rather than the cubic scaling of traditional methods). ONETEP\edit{\cite{Skylaris2005a,Prentice2020}} is one such code, notable for its equivalence to plane-wave approaches due to the \emph{in situ} optimization of its basis (a set of local Wannier-like orbitals). Its ability to routinely perform DFT calculations on systems containing thousands of atoms allows more detailed study of nanostructures\cite{Heiss2013a,Todorova2013a}, defects\cite{Hine2010a, Corsetti2011a}, and biological systems\cite{Dziedzic2013a,Lever2014,Fokas2017,Cole2016a}.

That said, DFT is not without its shortcomings. Many of these stem from its approximate treatment of exchange and correlation via an exchange-correlation (XC) functional. These shortcomings become especially evident in ``strongly-correlated'' systems, which typically contain transition element or rare-earth atoms whose $3d$- or $4f$-electron shells are partially filled. Electrons in these shells are in especially close proximity with one another, and their interaction is too pronounced to be adequately described by DFT, which can provide even qualitatively incorrect descriptions of the electronic structure. For example, DFT often yields magnetic moments inconsistent with experiment,\cite{Tran2006a} 
predicting some insulators to be metallic,\edit{\cite{Rodl2009a,Anisimov1997b}} 
and yielding equilibrium volumes dramatically different to experiment.\cite{Becker1996} 
DFT also fails to capture important dynamic properties that are enhanced by strong correlation, such as satellite peaks in photoemission spectra.\cite{McMahan2003,Lichtenstein2001}

These cases motivate the need for more accurate theories. One such approach is dynamical mean field theory (DMFT),\edit{\cite{Georges1992a,Georges1996a,Kotliar2006,Held2007,Gull2011}} a Green's function method that maps the lattice electron problem onto a single-impurity Anderson model with a self-consistency condition. Local quantum fluctuations are fully taken into account, allowing DMFT to capture complex electronic behavior such as the intermediate three-peak states of the Mott transition, the transfer of spectral weight, and the finite lifetime of excitations\cite{Georges1996a}. Furthermore, it is possible to embed DMFT within a DFT framework, whereby only atoms with strongly-correlated electrons are treated at the DMFT level, while the rest of the system can be treated at the DFT level.\edit{\cite{Anisimov1997b}} This is critical, as DMFT alone is prohibitively expensive for studying most realistic computational models.

In the past decade, numerous codes have been written to add DMFT functionality to existing DFT packages. These include EDMFTF\cite{Haule2010, HauleWebsite} and DFTTools\cite{Aichhorn2016} on top of Wien2K\cite{Wien2KWebsite}, EDMFTF\cite{HauleWebsite} on top of VASP \cite{Kresse1996, Kresse1996b, VASPWebsite}, DCore\cite{DCoreWebsite} on top of Quantum Espresso \cite{Giannozzi2009} and OpenMX\cite{Ozaki2005a, OpenMXWebsite}, TOSCAM\cite{Plekhanov2018} on top of CASTEP\cite{Clark2005,CASTEPWebsite}, Amulet\cite{AmuletWebsite} on top of Quantum Espresso \cite{Giannozzi2009} and Elk \cite{ElkWebsite}, and ComDMFT\cite{Choi2018} on top of FlapwMBPT\cite{FlapwMBPTWebsite, Kutepov2016}. Many of these make use of stand-alone libraries such as TRIQS\cite{Parcollet2015}, ALPS\cite{Gaenko2017}, iQIST \cite{Huang2015}, or W2dynamics\cite{Wallerberger2018}. This paper introduces the implementation of TOSCAM (A TOolbox for Strongly Correlated Approaches to Molecules) on top of ONETEP, a linear-scaling DFT code. In contrast to the packages mentioned above, this approach uniquely enables us to perform DMFT calculations on large and aperiodic systems such as nanoparticles and metalloproteins.

This code has already seen success: it has been used to explain the insulating $M_1$ phase of vanadium dioxide,\cite{Weber2012a} to demonstrate the importance of Hund's coupling in the binding energetics of myoglobin,\cite{Weber2013,Weber2014a} and to reveal the super-exchange mechanism in the di-Cu oxo-bridge of hemocyanin and tyrosinase.\cite{Al-Badri2018}. But until now it has not been available to the scientific community at large. The DMFT module in ONETEP has been included in version 5.0, and TOSCAM is being released at \texttt{<github link to accompany publication>}. This paper presents an overview of this methodology, its implementation, and an example of its application. 

\section{Theory}
\subsection{The ONETEP framework}
\begin{figure}[!t]
   \footnotesize
   \centering
   \begin{tikzpicture}[font=\small]
      \begin{scope}[on background layer]
         \node [dummy](dummy){};
         \node[dummy,xshift=3cm,yshift=0cm](dummy2) at (dummy.east){};
         \node[dummy,xshift=4.5cm,yshift=0cm](dummy3) at (dummy.east){};
      \end{scope}

      \node[box1, left of = dummy, node distance=0cm, text width=3cm,align=center](guess){%
         Guess ${K}^{\alpha \beta}$ and $\{\phi_\alpha\}$
      };
      \node[box1, below=0.5cm of guess, text width=3cm,align=center](improveK){%
         Improve guess
         of ${K}{\alpha \beta}$
      };
      \node[box1, below=0.5cm of improveK, text width=3cm,align=center](converge?K){Converged with respect to ${K}^{\alpha \beta}$?};
      \node[box0, below=0.35cm of converge?K, text width=0.7cm,align=center](YesK){Yes};
      \node[box0, below of = dummy2, node distance=2.25cm, text width=0.7cm,align=center](NoK){No};
   
      \node[box1, below=0.35cm of YesK, node distance=5.1cm, text width=3cm,align=center](improveNGWFs){%
         Improve guess
         of $\{\phi_\alpha\}$
      };
      \node[box1, below=0.5cm of improveNGWFs, node distance=6.6cm, text width=3cm,align=center](converge?NGWFs){Converged with respect to $\{\phi_\alpha\}$?};
      \node[box0, below=0.35 of converge?NGWFs, node distance=7.6cm, text width=0.7cm,align=center](YesNGWFs){Yes};
      \node[box0, below of = dummy3, node distance=4.05cm, text width=0.7cm,align=center](NoNGWFs){No};
      \node[box1, below=0.35 of YesNGWFs, node distance=8.45cm, text width=3cm,align=center](Solfound){Solution found};
      \path [line] (guess) -- (improveK);
      \path [line] (improveK) -- (converge?K);
      \path [draw, ultra thick, -] (converge?K) -| (NoK);
      \path [line] (NoK) |- (improveK);
      \path [draw, ultra thick, -] (converge?K) -- (YesK);
      \path [line] (YesK) -- (improveNGWFs);
      \path [line] (improveNGWFs) -- (converge?NGWFs);
      \path [draw, ultra thick, -] (converge?NGWFs) -| (NoNGWFs);
      \path [line] (NoNGWFs) |- (improveK);
      \path [draw, ultra thick, -] (converge?NGWFs) -- (YesNGWFs);
      \path [line] (YesNGWFs) -- (Solfound);
   
   \end{tikzpicture}
   \caption{Process by which ONETEP finds a self-consistent ground-state solution for $K^{\alpha \beta}$ and $\{\phi_\alpha\}$.}
   \label{fig:ONETEP_flowchart}
\end{figure}
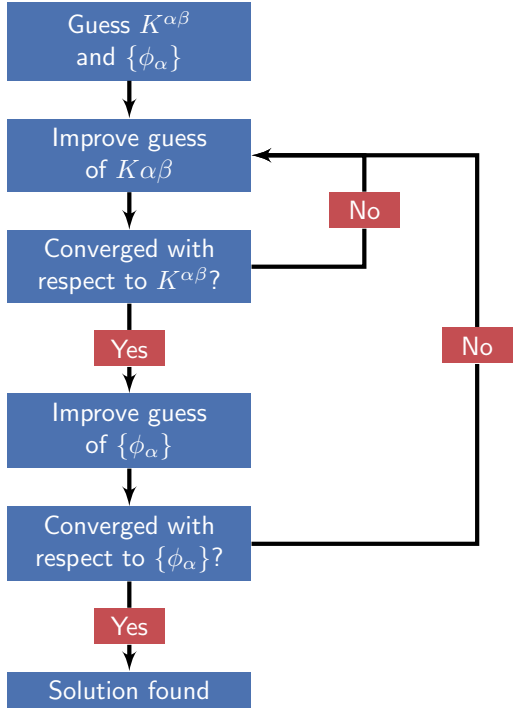

In the ONETEP implementation of linear-scaling DFT, we work with the single-particle density-matrix:
\begin{equation}
\rho(\mathbf{r},\mathbf{r}') = \sum_{\alpha, \beta} \phi_\alpha(\mathrm{r})K^{\alpha\beta} \phi_{\beta}(\mathrm{r}'),
\end{equation}
where $\{\phi_{\alpha}\}$ are a set of localized non-orthogonal generalized Wannier functions (NGWFs) and $K^{\alpha \beta}$ is the density kernel. These orbitals are variationally optimized \emph{in situ} during the energy-minimization carried out as part of the DFT calculation.\cite{Skylaris2002a}. For the purposes of this optimization, the NGWFs are in turn expanded in terms of a systematic basis of psinc functions\cite{Mostofi2003a} --- systematic, in the sense that the size of this basis is determined solely by a scalar parameter (a plane-wave kinetic energy cutoff determining the grid spacing) that can be increased until convergence is reached. In this scheme, a DFT calculation does not involve cyclically calculating the Kohn-Sham density and potential, but instead involves the direct minimization of the DFT energy with respect to both the density kernel and the NGWF expansion coefficients (Figure~\ref{fig:ONETEP_flowchart}). Due to the fact that the NGWFs are localized, the associated matrix algebra is sparse. Meanwhile, because the NGWFs are optimized \emph{in situ}, the calculations are not prone to basis set incompleteness or superposition error\cite{Haynes2006}, while at the same time permitting a relatively small number of basis functions.\cite{Skylaris2002a}

A fully converged energy minimization yields the Kohn-Sham Hamiltonian $H_{\alpha\beta}$ in the NGWF representation, and from this, related properties such as orbital energies, electronic and spin densities, densities of states etc can be obtained. For many systems this will provide an adequate description of their electronic structure. However, in cases where we have strong electronic correlation, this Hamiltonian must be improved upon. This is the job of dynamical mean field theory.

A DMFT calculation involves the self-consistent calculation of the Green's function ${G}^{\alpha\beta}(\omega)$ ($\omega$ here may be $\omega + i\eta$ or $i\omega_n$ if operating in the finite-temperature Matsubara representation) and the self-energy ${\Sigma}_{\alpha\beta}$, which are related via
\begin{equation}
G^{\alpha \beta}(\omega) = {\left[(\omega + \mu)S - H - \Sigma(\omega)\right]_{\alpha\beta}}^{-1}
\label{eqn:G_tot}
\end{equation}
where $\mu$ is the chemical potential (fixed at the mid-point of the energies of the highest occupied and lowest unoccupied KS orbitals), and ${S}_{\alpha\beta}$ is the NGWF overlap matrix (that is, ${S}_{\alpha\beta} = \braket{\phi_\alpha}{\phi_\beta}$), which is non-diagonal.

Treating most physical systems at the DMFT level would usually be prohibitively expensive. The DFT\,+\,DMFT scheme takes advantage of the fact that strong electronic correlation is often confined to identifiable localized subspaces (for instance, the $3d$ orbitals of a transition metal atom), with the remainder of the system having a delocalized, free-electron character. In such systems, the correlated subspaces can be treated at the DMFT level, while DFT alone should be sufficient everywhere else.

Correlated subspaces are typically defined via a set of local, fixed, atom-centered, spin-independent, and orthogonal orbitals $\{\tilde \varphi^I_m\}$. (Here, $I$ is the atom index and $m$ is an orbital index.) In ONETEP, these are defined using \emph{pseudoatomic orbitals} $\{\varphi^I_m\}$, the Kohn-Sham solutions to the isolated pseudopotential of the correlated atom.\cite{Sankey1989a,Artacho1999a,Ruiz-Serrano2012a} The two are related via $\ket{\tilde \varphi^I_m} = \ket{\phi_\alpha}\braket{\phi^\alpha}{\varphi^I_m}$. (Note that the pseudoatomic orbitals do not necessarily reside in the subspace spanned by the NGWFs.)

\subsection{The Anderson impurity model}

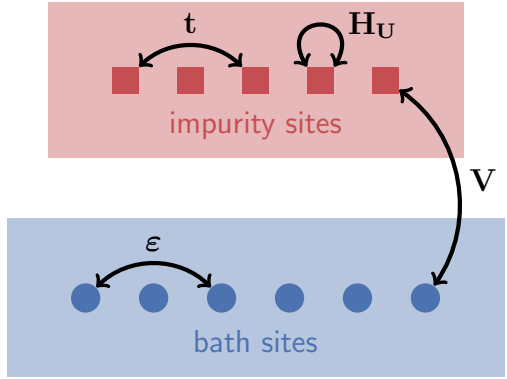
\begin{figure}[!t]
\centering
\newlength{\figscale}
\setlength{\figscale}{0.5cm}

\begin{center}
\begin{tikzpicture}[font=\sffamily, ]
   \begin{scope}[on background layer]
      \node [dummy](dummy){};
   \end{scope}

\tikzstyle{square} = [rectangle, fill=seabornred, %
   text centered, text=black, font=\tiny, minimum size=0.7\figscale, minimum width=0.7\figscale]
\tikzstyle{mycircle} = [circle, fill=seabornblue, %
   text centered, minimum size=0.7\figscale, font=\footnotesize]

   \node[square, below of = dummy        ](impurity3){};
   \node[square, left= \figscale of impurity3](impurity2){};
   \node[square, left= \figscale of impurity2](impurity1){};
   \node[square, right=\figscale of impurity3](impurity4){};
   \node[square, right=\figscale of impurity4](impurity5){};

   \coordinate (dummy2) at ($(dummy.south) + (0.0cm,-7.5\figscale)$);
   \node[mycircle, left= 0.5\figscale of dummy2                            ](bath3){};
   \node[mycircle, right=0.5\figscale of dummy2                            ](bath4){};
   \node[mycircle, left= \figscale    of bath3                             ](bath2){};
   \node[mycircle, left= \figscale    of bath2                             ](bath1){};
   \node[mycircle, right=\figscale    of bath4                             ](bath5){};
   \node[mycircle, right=\figscale    of bath5                             ](bath6){};

 \draw[<->, ultra thick] (bath1)     to [bend left =45] node[above]{$\boldsymbol{\varepsilon}$} (bath3);
 \draw[<->, ultra thick] (bath6)     to [bend right=45] node[right]{$\bV$}                   (impurity5);
 \draw[<->, ultra thick] (impurity1) to [bend left =45] node[above]{$\bt$}                                    (impurity3);
 \draw[<->, ultra thick] (impurity4.north west) to [out=120, in=60, looseness=6] node[right]{\hspace{0.2cm}$\mathbf{H_U}$} (impurity4.north east);

\begin{pgfonlayer}{background}
   \node[box0, fit= (impurity1) (impurity2) (impurity3) (impurity4) (impurity5), 
         draw=none, fill opacity=0.4, inner sep=1.7\figscale](impurities){};
   \node[box1, fit= (bath1) (bath2) (bath3) (bath4) (bath5) (bath6),
         draw=none, fill opacity=0.4, inner sep=1.7\figscale](bath){};
   \node[dummy, below=-1.55\figscale of impurities, seabornred ]{impurity sites};
   \node[dummy, below=-1.55\figscale of bath,       seabornblue]{bath sites};
\end{pgfonlayer}
\end{tikzpicture}
\end{center}
\caption{Schematic diagram of an Anderson impurity model, showing the impurity sites (orange squares), bath sites (purple circles) and the interaction parameters.}
\label{fig_cartoon_aim}
\end{figure}

In order to efficiently find a self-consistent solution to equation \ref{eqn:G_tot}, DMFT relies on mapping correlated subspaces to auxiliary Anderson impurity models (AIMs). The AIM is a simplified Hamiltonian that describes the interaction of a number of sites (known as impurity sites) with a bath of additional electronic levels:
\begin{equation}
\hat H =
\underbrace{\sum_{ij\sigma} (\varepsilon_{ij}-\mu) \hat c^\dag_{i\sigma} \hat c_{j\sigma}}_{\hat H_\text{bath}}
+ \underbrace{\sum_{im\sigma} \left(V_{m i}\hat f^\dag_{m\sigma} \hat c_{i\sigma}
+ h.c.\right)}_{\hat H_\text{mix}}
+ \underbrace{
\sum_{mm'\sigma} (t_{mm'}-\mu)\hat f^\dag_{m\sigma}\hat f_{m'\sigma} + \hat H_{U}
}_{\hat H_\text{loc}}
\label{eqn:AIM_hamiltonian}
\end{equation}
where $\hat H_\text{bath}$ describes the non-correlated behavior of the bath (parameterized by the hopping matrix $\varepsilon_{ij}$), $\hat H_\text{loc}$ the impurity (parameterized by the impurity hopping $t_{mm'}$ and the interaction Hamiltonian $\hat H_U$), and $\hat H_\text{mix}$ the coupling between the two (parameterized by $V_{mi}$). This Hamiltonian is depicted pictorially in Fig.~\ref{fig_cartoon_aim}. The bath and impurity sites have a shared chemical potential $\mu$, and $\hat c$/$\hat f$ are the annihilation operators for the bath/impurity. The convention throughout will be that Greek indices correspond to NGWFs, $m$ and $m'$ to Hubbard subspaces and their corresponding impurity sites, and Latin indices to bath sites. $\sigma$ is the spin index.

The non-interacting Anderson model (\emph{i.e.} $H_U = 0$) has the Green's function
\begin{equation}
G^0_\text{tot}(\omega) = \frac{1}{\omega+\mu-T}
\end{equation}
where the full hopping matrix is of the block matrix form
\begin{equation}
T = 
\begin{pmatrix}
t      & V \\
V^\dag & \varepsilon \\
\end{pmatrix}.
\end{equation}
It follows that the (non-interacting) impurity Green's function --- that is, the top-left-hand block of $G^0_\mathrm{tot}(\omega)$ --- simplifies to
\begin{equation}
{{G}_\text{imp}^0}(\omega)^{-1}
= \omega +\mu - t - 
\Delta_\text{imp}(\omega)
,
\label{eqn:DMFT_AIM_Gimp_nonint}
\end{equation}
where
\begin{equation*}
{\Delta_\text{imp}}_{mm'}(\omega) = V_{mi} \left(\frac{1}{\omega+\mu-{\varepsilon}}\right)_{ij} V^{\dag}_{jm'}
\label{eqn:DMFT_AIM_Delta}
\end{equation*}
is the so-called impurity \textit{hybridization function}. This quantity is of particular importance because it encapsulates all of the contributions of the bath sites to the physics of the impurity sites; the AIM impurity Green's function is given by
\begin{equation}
{G}_\text{imp}(\omega)^{-1} 
= {G}_\text{imp}^0(\omega)^{-1}
- {\Sigma}(\omega)
= \omega + \mu - t - {\Delta}_\text{imp}(\omega) - {\Sigma}_\mathrm{imp}(\omega)
\label{eqn:DMFT_AIM_Gimp}.
\end{equation}
%

\subsection{A DMFT calculation}
This subsection will walk through the steps in a standard DMFT calculation as performed in TOSCAM + ONETEP. It is important to note that DMFT typically invokes a mean field approximation across multiple correlated sites (hence \emph{dynamical ``mean field" theory}), an approach that only becomes exact in the limit of infinite coordination (or equivalently, dimensions). This is not the case in our following real-space approach, where correlated sites are treated as a (possibly multi-site) AIM. 

\subsubsection{Mapping physical systems to an impurity model}
DFT\,+\,DMFT utilizes an AIM as an auxiliary system: the AIM parameters $\{V_{m i}\}$, $\{\varepsilon_{ij}\}$, and $\{t_{mm'}\}$ are chosen such that the resulting model Hamiltonian reproduces the physics of the real system as closely as possible. This mapping proceeds as follows. Firstly, the Kohn-Sham Hamiltonian, an estimate of the system self-energy (zero is a reasonable starting point), and a total Green's function (obtained via equation \ref{eqn:G_tot}) are each projected onto the correlated subspaces. For instance, the local Green's function is given by
\begin{equation}
{{\tilde G}}^I_{mm'}(\omega) = {W}^I_{m\alpha}{G}^{\alpha\beta}(\omega) (W^I)^\dagger_{\beta m'}
\label{eqn:G_loc}
\end{equation}
where ${W}^I_{m\alpha} = \braket{\varphi_m^I}{\phi_\alpha}$ is the overlap of the NGWFs and the Hubbard projectors. In a similar manner one can obtain the projected self energy $\tilde \Sigma^I(\omega)$ and the projected Kohn-Sham Hamiltonian $\tilde H^I$.

We can determine the appropriate impurity hopping parameters $t_{mm'}$ for the AIM by comparing the AIM impurity Green's function $G_\text{imp}(\omega)$ and the local Green's function $G^I_{mm'}(\omega)$: if these are to match in the high=frequency limit to order $\mathcal{O}(1/\omega^2)$ then it follows that $t = \tilde O^I W^I S^{-1} H S^{-1} W^{I\dag} \tilde O^I$, where $\tilde O^I = (W^I S^{-1} W^{I\dag})^{-1}$ is the overlap matrix of the projectors ${\tilde \varphi_m}$. Meanwhile, in order to define $\{V_{mi}\}$ and $\{\varepsilon_{ij}\}$, we define the local hybridization function for our physical system
\begin{equation}
\tilde \Delta^I(\omega) = (\omega + \mu)\tilde O^I - (\tilde G^I)^{-1}(\omega) - \tilde \Sigma^I(\omega)- \tilde H^I
\label{eqn:Delta_loc}
\end{equation}
 which is analogous to the definition of the impurity hybridization function (equation~\ref{eqn:DMFT_AIM_Gimp}). We choose the impurity model bath parameters such that the AIM hybridization function matches this local hybridization function as closely as possible. This is done by minimizing the distance function

\begin{equation}
d(V, \varepsilon) = \sum_{\omega < \omega_c}\frac{1}{\omega^\gamma} \left| \Delta_\text{imp}(\omega) - \tilde \Delta^I(\omega) \right|^2
\label{eqn:DMFT_distance_function}
\end{equation}
using a conjugate gradient (CG), Broyden-Fletcher-Goldfarb-Shanno (BFGS), or similar minimization algorithm.  Here, $\omega_c$ is a cutoff frequency and $\gamma$ is a user-specified parameter that can allow for preferential weighting of agreement at low frequencies.

In order to complete the construction of the auxiliary AIM Hamiltonian we choose $H_U$ to be of the Slater-Kanamori form\cite{Slater1936a, Kanamori1959a}
\begin{align}
\hat H_U = U\sum_m \hat n_{m\uparrow}\hat n_{m\downarrow}
+ \left(U'-\frac{J}{2}\right) \sum_{m>m'} \hat n_m \hat n_{m'} \nline
- J \sum_{m>m'} (2\hat \bS_m\hat\bS_{m'}+
\hat f^\dag_{m\uparrow}\hat f^\dag_{m\downarrow}
\hat f_{m'\uparrow}\hat f_{m'\downarrow}).
\end{align}
This Hamiltonian is well-suited to capturing multiplet properties of low energy states.\cite{Imada1998} Its first term describes intra-orbital Coulomb repulsion. The second describes the inter-orbital repulsion, with $U' = U - 2J$ further renormalized by the Hund's coupling to ensure the rotational invariance of the Hamiltonian. The third and final term captures the Hund's exchange coupling; $\hat \bS_m$ is the spin of orbital $m$, given by $(\hat \bS_m)_i = \frac{1}{2}\sum_{\sigma\sigma'}\hat f^\dag_{m\sigma}(\bs_i)_{\sigma\sigma'}\hat f_{m\sigma'}$ via the Pauli spin matrices $\{\bs_i\}$. The Hubbard parameter $U$ and Hund's coupling $J$ are user-specified parameters that in principle could be obtained via linear response \cite{Cococcioni2005a} but are often chosen empirically or treated as variational parameters.

Now that we have defined $\varepsilon$, $V$, $t$, and $H_U$, the mapping of a real system to an auxiliary AIM is complete. In theory, this mapping can be exact: as long as $\Delta_\text{imp}(\omega)$ and $\tilde \Delta^I(\omega)$ match exactly, $G_\text{imp}(\omega)$ and $\tilde G^I(\omega)$ will also. Getting this mapping right is therefore of the utmost importance.

\subsubsection{Solving the AIM}
Having constructed the AIM Hamiltonian $H_\text{AIM}$, the next step is to calculate the Green's function of the AIM (known as the \emph{impurity Green's function}):
\begin{align}
{G_\text{imp}}_{mm'}(\omega)
&= \int_{-\infty}^\infty e^{i\omega t} {G_\text{imp}}_{mm'}(t) \, dt \nline
&= - i \int_{0}^\infty e^{i\omega t} \langle {e^{i\hat H t} \hat c_m e^{-i\hat H t}, \hat c^\dag_{m'}} \rangle \, dt \nline
&= - i \left(
    \left\langle \hat c_m \int_{0}^\infty e^{i(\omega -(\hat H - E_0))t} \, dt \, \hat c^\dag_{m'} \right\rangle
  + \left\langle \hat c^\dag_{m'} \int_{0}^\infty e^{i(\omega +(\hat H - E_0))t} \, dt \, \hat c_m \right\rangle
\right)\nline
&= \left\langle \hat c_m \frac{1}{\omega -(\hat H - E_0)} \hat c^\dag_{m'} \right\rangle
  + \left\langle \hat c^\dag_{m'} \frac{1}{\omega +(\hat H - E_0)} \hat c_m \right\rangle 
\label{eqn:DMFT_Lanczos_Gimp}
\end{align}
where $\langle~\bullet~\rangle$ is the thermodynamic average, which at zero temperature becomes $\bra{\psi_0}~\bullet~\ket{\psi_0}$.

Resolving equation~\ref{eqn:DMFT_Lanczos_Gimp} is highly expensive, and becomes one of the most substantial computational barriers in a DMFT calculation. If there are $m$ bath sites and $n$ impurity orbitals, the Hilbert space of this problem scales as $4^{m+n}$. (For a system containing a single transition metal there will be five impurity orbitals --- one for each $3d$ orbital --- and then typically six to eight bath sites.) This is far larger than any of the other matrix inversions that we need to calculate during the DMFT loop (for instance, $G^{\alpha\beta}$ is only as large as the number of Kohn-Sham orbitals, which in turn will be of the order of the number of electrons in the physical system --- typically several thousand at most). There are a number of approaches for obtaining  $G_\text{imp}$, such as exact diagonalization (ED) and continuous time Monte Carlo algorithms. The calculations in this work employ ED via the Lanczos algorithm to evaluate equation~\ref{eqn:DMFT_AIM_Gimp}, a process which is explained in detail in the Supplementary Information.

Given a solution $G_{\text{imp}}$ (obtained via ED or otherwise), the impurity self-energy can then be obtained via
\begin{equation}
\Sigma_\text{imp}(\omega)=[G^0_\text{imp}]^{-1}(\omega)-G_\text{imp}^{-1}(\omega)
\label{eqn:DMFT_Sigma_imp}
\end{equation}
where the non-interacting impurity Green's function is given by equation \ref{eqn:DMFT_AIM_Gimp_nonint}. Note that this operation is far less expensive than equation~\ref{eqn:DMFT_AIM_Gimp} because these matrices are only $m \times m$ in size.

\subsubsection{Upfolding and double-counting}
Having obtained the impurity Green's function $\Sigma_\text{imp}^I$ for each AIM, the final step is to upfold this result to the complete physical system. Since the original DFT solution already contains the influence of the Coulomb interaction to some degree, double-counting becomes an issue. A popular form of the correction is
\begin{equation}
E_\mathrm{DC} = \frac{U^\mathrm{av}}{2}n\left(n - 1 \right)-\frac{J}{2}\sum_\sigma n_\sigma(n_\sigma-1)
\label{eqn:DMFT_Edc}
\end{equation}
where $n$ is the total occupancy of the subspace, $n^\sigma$ is the occupancy of the subspace for the spin channel $\sigma$, and
\begin{equation}
U^\mathrm{av} = \frac{U + 2(N-1)U'}{2N-1}
\end{equation}
with $N$ being the number of orbitals spanning the correlated subspace (and recall that $U' = U - 2J$).\cite{Imada1998} This double-counting is derived by attempting to subtract the DFT contributions in an average way; $U^{av}$ is the average of the intra- and inter-orbital Coulomb parameters.

The self-energy is upfolded to the NGWF basis via 
\begin{equation}
\Sigma_{\alpha\beta} = \sum_I {W^I}^{\dag}_{\alpha m} ({\Sigma^I}^{mm'} - E_\mathrm{DC}\tilde O^{mm'})W^I_{m'\beta}
\label{eqn:DMFT_Sigma_upfolded}
\end{equation}
--- and with that, we are back where we started, having generated a new estimate of the self-energy $\Sigma_{\alpha\beta}$ for the full system.

To summarize, the scheme is as follows:
\begin{enumerate}
\item perform a DFT calculation to construct the system Hamiltonian
\item initialize the self-energy as $\Sigma_{\alpha \beta}(\omega)=0$
\item obtain the Green's function for the full system (equation~\ref{eqn:G_tot})
\label{item:dmft_obtain_total_GF}
\item project the total Green's function and self energy onto the $I$\textsuperscript{th} Hubbard subspace to obtain the corresponding local quantities (equation \ref{eqn:G_loc})
\item calculate the local hybridization function (equation \ref{eqn:Delta_loc})
%
\item find the bath parameters $\varepsilon_{ij}$ and $V_{mi}$ such that the AIM hybridization function (equation~\ref{eqn:DMFT_AIM_Delta})
matches the local hybridization function found above
\item explicitly solve the AIM Hamiltonian to obtain the impurity Green's function (equation~\ref{eqn:DMFT_Lanczos_Gimp})
\item update the impurity self-interaction (equation~\ref{eqn:DMFT_Sigma_imp})
\item upfold the self-energies from each correlated subspace to obtain the total self-interaction (equation~\ref{eqn:DMFT_Sigma_upfolded})
\label{item:dmft_steps_upfold}
\end{enumerate}

Note that if we only have one correlated site in our system, this mapping is exact, and the local lattice Green's function at step \ref{item:dmft_steps_upfold} will already match the impurity Green's function.

This is not the case for bulk systems. There, the mean field approximation that we adopt means that the self-energy of a correlated site is also inherited by the ``bath" \emph{i.e.} one would typically solve a single Anderson impurity problem but then in equation \ref{eqn:DMFT_Sigma_upfolded}, the index $I$ would run over all correlated sites. This means that after step \ref{item:dmft_steps_upfold} we must return to step \ref{item:dmft_obtain_total_GF}, and repeat this loop until the local lattice and impurity Green's functions match.

Once the calculation is converged, we can extract system properties from the Green's function. \edit{For example, DFT+DMFT total energies can be calculated via
\begin{equation}
    E = E_\mathrm{DFT}[\rho] - \sum_{i\sigma} \varepsilon_{i\sigma}  + \Trace{H_\mathrm{KS}G} + \frac{1}{2}\Trace{\Sigma G} - E_\mathrm{DC} + E_\mathrm{nuc-nuc}
\end{equation}
where the first term is the DFT energy functional, $\varepsilon_{i\sigma}$ are the Kohn-Sham eigenenergies, $H_{KS}$ is the Kohn-Sham Hamiltonian, $G$ is the full Green's function, the fourth term is the DMFT contribution to the potential energy $\langle \hat H_U \rangle$ calculated via the Migdal-Galitskii formula, the fifth term is the double-counting correction (equation \ref{eqn:DMFT_Edc}), and the final term is due to the interaction between atomic cores (that is, both the nuclei and the pseudized electrons).\cite{Kotliar2006,Amadon2006,Haule2015} When calculating the DMFT potential energy contribution, we use the high-frequency expansion technique of Pourovskii \emph{et al.}\cite{Pourovskii2007} For an example of ONETEP+TOSCAM DFT+DMFT energy calculations, see Refs.~\onlinecite{Weber2013} and ~\onlinecite{Weber2014a}, where some of us presented a detailed study of heme and myoglobin binding to CO and other small molecules.}

\edit{Other quantities that can be extracted from the Green's function include the density of states and the optical absorption.} One can also apply standard ONETEP analysis techniques to the electron density (such as natural bonding orbital analysis). These techniques will be demonstrated in Section~\ref{sec:application}.

\subsection{Extensions}
There are several possible extensions to the theory described thus far. These are not essential but often useful.

\subsubsection{Enlarged AIM via cluster perturbation theory}
If an AIM has too few bath sites at its disposal, it will be insufficiently flexible to fit a given local hybridization function. The brute-force approach would be to increase the number of bath sites, but in practice the number of bath sites is severely limited due to the exponential growth of Hilbert space with respect to the total number of sites (bath and impurity) of the AIM. To overcome this barrier, a secondary set of bath levels are coupled to the primary bath levels via cluster perturbation theory. By indirectly including these sites, the AIM system acquires extra flexibility without expanding the Hilbert space, resulting in a dramatic drop in the distance function. For more details, see Ref.~\onlinecite{Weber2012b}.

\subsubsection{Self-consistency}

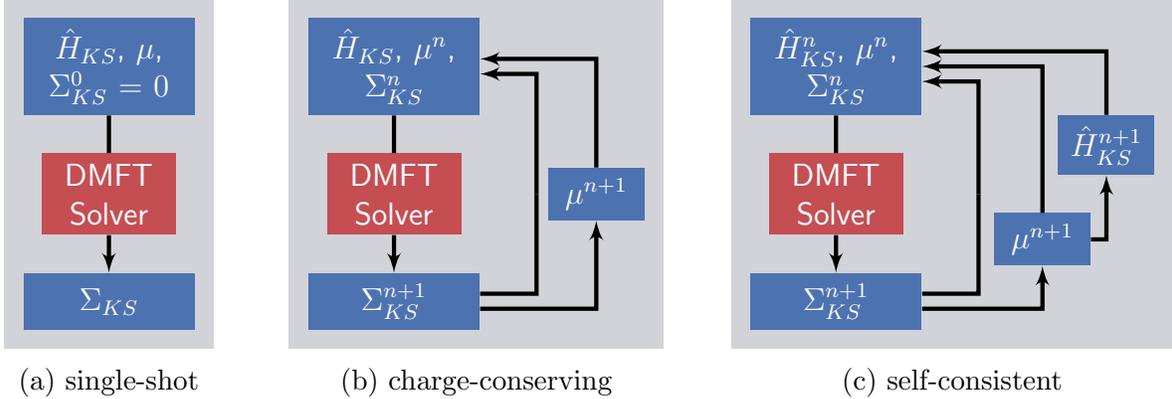
\begin{figure}[!t]
   \centering
   \begin{subfigure}[t]{0.225\textwidth}
      \centering

\begin{tikzpicture}[font=\tiny]
   \begin{scope}[on background layer]
      \node [dummy](dummy){};
   \end{scope}

   \node[box1, below of = dummy, text width=2cm,align=center](input){%
      $\hat H_{KS}$, $\mu$, $\Sigma^0_{KS}=0$ };

   \node[box0, below=0.5 of input, text width=1.5cm,align=center](solver){DMFT Solver};

   \node[box1, below=0.5cm of solver, text width=2cm,align=center](output){%
      $\Sigma_{KS}$\vphantom{$\Sigma^{n+1}_{KS}$}};
   
   \path [draw, ultra thick, -] (input) -- (solver);
   \path [line] (solver) -- (output);
\begin{pgfonlayer}{background}
   \node[box2, fit= (input) (solver) (output),
         draw=none, inner sep=0.25cm](background){};
\end{pgfonlayer}

\end{tikzpicture}
   \caption{single-shot}
   \label{fig:DMFT_scheme_singleshot}
   \end{subfigure}
   \begin{subfigure}[t]{0.35\textwidth}
      \centering

\begin{tikzpicture}[font=\tiny]
   \begin{scope}[on background layer]
      \node [dummy](dummy){};
   \end{scope}

   \node[box1, below of = dummy, text width=2cm,align=center](input){%
      $\hat H_{KS}$, $\mu^n$, \\ $\Sigma^n_{KS}$ };
   
   \node[box0, below=0.5 of input, text width=1.5cm,align=center](solver){DMFT Solver};

   \node[box1, below=0.5cm of solver, text width=2cm,align=center](output){%
      $\Sigma^{n+1}_{KS}$};
   \coordinate (dummy2) at ($(solver.east) + (1cm,0cm)$);
   \node[box1, right=1.15cm of solver, text width=1cm,align=center](muupdate){$\mu^{n+1}$};

   \path [draw, ultra thick, -] (input) -- (solver);
   \path [line] (solver) -- (output);
   \path [draw, ultra thick, -] ([yshift=0.1cm]output.east) -| (dummy2);
   \path [line] (dummy2) |- ([yshift=-0.1cm]input.east);
   \path [line] ([yshift=-0.1cm]output.east) -| (muupdate);
   \path [line] (muupdate) |- ([yshift=0.1cm]input.east);
\begin{pgfonlayer}{background}
   \node[box2, fit= (input) (solver) (output) (muupdate),
         draw=none, inner sep=0.25cm](background){};
\end{pgfonlayer}

\end{tikzpicture}
   \caption{charge-conserving}
   \label{fig:DMFT_scheme_chargesc}
   \end{subfigure}
   \begin{subfigure}[t]{0.4\textwidth}
      \centering
\begin{tikzpicture}[font=\tiny]
   \begin{scope}[on background layer]
      \node [dummy](dummy){};
   \end{scope}

   \node[box1, below of = dummy, text width=2cm,align=center](input){%
      $\hat H_{KS}^n$, $\mu^n$, \\ $\Sigma^n_{KS}$ };
   
   \node[box0, below=0.5 of input, text width=1.5cm,align=center](solver){DMFT Solver};

   \node[box1, below=0.5cm of solver, text width=2cm,align=center](output){%
      $\Sigma^{n+1}_{KS}$};
   \coordinate (dummy2) at ($(solver.east) + (1cm,0cm)$);
   \coordinate (dummy3) at ($(solver.east) + (1.85cm,0cm)$);
   \coordinate (dummy4) at ($(solver.east) + (2.7cm,0cm)$);
   \node[box1, below=0.25cm of dummy3, text width=1cm,align=center](muupdate){$\mu^{n+1}$};
   \node[box1, above=0.25cm of dummy4, text width=1cm,align=center](Hupdate){$\hat H_{KS}^{n+1}$};

   \path [draw, ultra thick, -] (input) -- (solver);
   \path [line] (solver) -- (output);
   \path [draw, ultra thick, -] ([yshift=0.1cm]output.east) -| (dummy2);
   \path [line] (dummy2) |- ([yshift=-0.2cm]input.east);
   \path [line] ([yshift=-0.1cm]output.east) -| (muupdate);
   \path [line] (muupdate) |- (input.east);
   \path [line] (muupdate.east) -| (Hupdate);
   \path [line] (Hupdate) |- ([yshift=0.2cm]input.east);
\begin{pgfonlayer}{background}
   \node[box2, fit= (input) (solver) (output) (muupdate) (Hupdate),
         draw=none, inner sep=0.25cm](background){};
\end{pgfonlayer}

\end{tikzpicture}
   \caption{self-consistent}
   \label{fig:DMFT_scheme_fullysc}
   \end{subfigure}
   \caption[Three DMFT schemes: single-shot, charge-conserving, and self-consistent]{The three DMFT schemes, in increasing order of complexity.}
   \label{fig:DMFT_schemes}
\end{figure}

For a system with a single correlated site, there is no feedback from the self energy to the hybridization function, and --- provided the AIM is sufficiently representative --- the DMFT algorithm will converge in a single step. 
In this case the algorithm is not a mean-field approximation, but exact. This scheme is shown in Fig.~\ref{fig:DMFT_scheme_singleshot}.


However, there are a number of reasons why we may not be content with the resulting solution. Firstly, the total number of electrons in the system is related to the total retarded Green's function via
\begin{equation}
N = \int d\omega\, \rho^{\alpha\beta}(\omega) S_{\alpha\beta}; \qquad \rho^{\alpha\beta}(\omega) = \frac{1}{2\pi i}\left(G^{\alpha\beta}(\omega) - {G^{\alpha\beta}}^\dag(\omega) \right),
\end{equation}
where $\rho^{\alpha\beta}(\omega)$ is the basis-resolved DMFT density matrix. 
There is no reason \emph{a priori} why the Green's function, updated via the DMFT loop, should yield the same number of electrons as we started with --- in fact, this is almost never the case. For this reason, charge conservation can optionally be enforced by adjusting $\mu$ so that $\int_{-\infty}^{\mu} \rho(\omega) = N$. This update is performed during each DMFT cycle, which means that our total Green's function (now adjusted by our altered $\mu$) will not necessarily be consistent with the self energy --- and consequently more than one DMFT loop will likely be required to iterate to self-consistency (Fig.~\ref{fig:DMFT_scheme_chargesc}). We will refer to this as ``charge-conserving" DMFT.

Finally, in the DFT formalism, the Hamiltonian is a functional of the density. It could be argued that if we are to be fully self-consistent, if ever the density changes the Hamiltonian should be updated accordingly. In this scheme, one iterates until $\Sigma$, $H$, and $\mu$ all converge (Fig.~\ref{fig:DMFT_scheme_fullysc}). This we will refer to as ``fully self-consistent" DMFT. We use Pulay mixing\cite{Pulay1980,Pulay1982} to update the Hamiltonian (via the density kernel) and the self-energy. Performing this double-loop naturally makes the calculations much more expensive, but they remain feasible. This approach was taken in Refs.~\onlinecite{Pourovskii2007}\nocite{Park2014}--\onlinecite{Bhandary2016}, for example.

\subsection{Practical implementation}
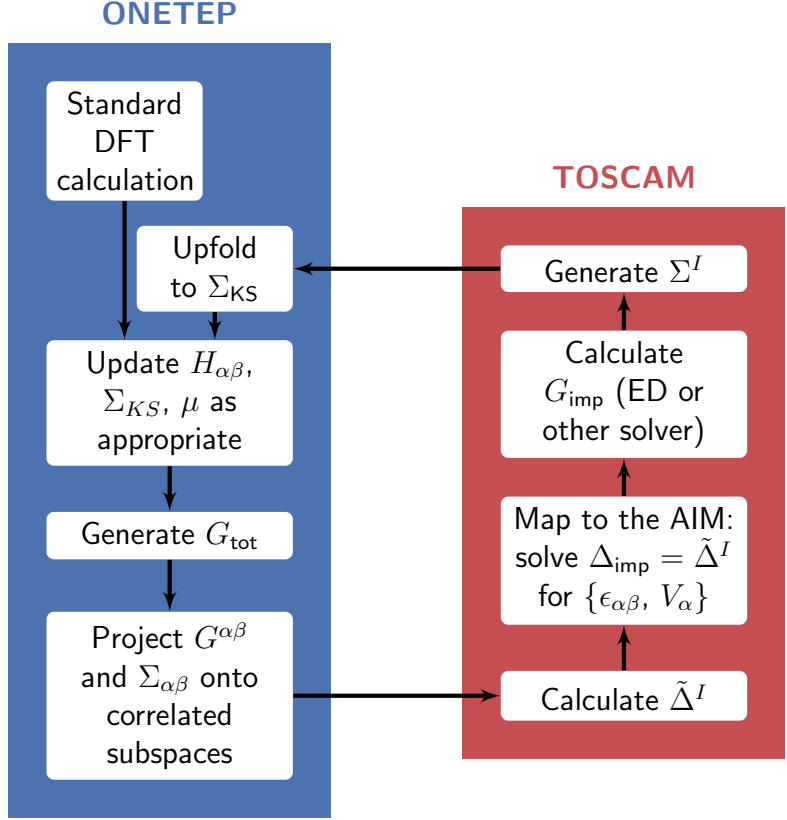
\begin{figure}[!t]
\centering
{
\begin{tikzpicture}[font=\tiny]
   \begin{scope}[on background layer]
      \node [dummy](dummy){};
      \node [dummy, left of = dummy, node distance=0.6cm](dummya){};
      \node [dummy, right of = dummy, node distance=0.6cm](dummyb){}; 
      \node [dummy, right of = dummy, node distance=6cm](dummy2){}; 
   \end{scope}

   \node[boxwhite, below of = dummya, node distance=0cm, text width=1.8cm,align=center](DFT){%
      Standard DFT \\ calculation};

   \node[boxwhite, below of = dummyb, node distance=1.7cm, text width=1.8cm,align=center](Upfold){Upfold to $\Sigma_\text{KS}$};

   \node[boxwhite, below=2.5cm of dummy, text width=3cm,align=center](Hks_sigma){%
      Update $H_{\alpha\beta}$, $\Sigma_{KS}$, $\mu$ as appropriate};
   
   \node[boxwhite, below=0.6cm of Hks_sigma, text width=3cm,align=center](Gtotal){
   Generate $G_\text{tot}$};

   \node[boxwhite, right=2.75cm of Upfold, node distance=3cm, text width=3cm,align=center](Sigmanew){%
      Generate $\Sigma^I$};
      
   \node[boxwhite, below=0.5cm of Sigmanew, text width=3cm,align=center](Gimp){%
      Calculate $G_\text{imp}$ (ED or other solver)};
      
   \node[boxwhite, below=0.5cm of Gimp, text width=3cm,align=center](SolveAIM){%
      Map to the AIM: solve $\Delta_\text{imp} = \tilde \Delta^I$ for \{$\epsilon_{\alpha\beta}$, $V_\alpha$\}};
   
   \node[boxwhite, below=0.6cm of SolveAIM, text width=3cm,align=center](hybfn){%
      Calculate $\tilde \Delta^I$};

   \node[boxwhite, left=2.75cm of hybfn, text width=3cm,align=center](Gloc){%
      Project $G^{\alpha\beta}$ and $\Sigma_{\alpha\beta}$ onto correlated subspaces};
   
   \path [line] (DFT) -- ([xshift=-0.6cm]Hks_sigma.north);
   \path [line] (Hks_sigma) -- (Gtotal);
   \path [line] (Gtotal) -- (Gloc);
   \path [line] (Gloc) -- (hybfn);
   \path [line] (hybfn) -- (SolveAIM);
   \path [line] (SolveAIM) -- (Gimp);
   \path [line] (Gimp) -- (Sigmanew);
   \path [line] (Sigmanew) -- (Upfold);
   \path [line] (Upfold) -- ([xshift=0.6cm]Hks_sigma.north);

\begin{pgfonlayer}{background}
   \node[box1, fit= (DFT) (Upfold) (Hks_sigma) (Gtotal) (Gloc),
         draw=none, fill opacity=1, inner sep=0.5cm](onetep){};
   \node[box0, fit= (Sigmanew) (Gimp) (SolveAIM) (hybfn), 
         draw=none, fill opacity=1, inner sep=0.5cm](toscam){};
   \node[dummy, above=0.1cm of onetep, color=seabornblue]{\normalsize \textbf{ONETEP}};
   \node[dummy, above=0.1cm of toscam, color=seabornred]{\normalsize \textbf{TOSCAM}};
\end{pgfonlayer}

\end{tikzpicture}
}
   \caption{A simplified DMFT loop, demonstrating which program (ONETEP or TOSCAM) is responsible for which step.}
   \label{fig:ONETEP_and_TOSCAM}
\end{figure}

In our implementation, ONETEP and TOSCAM are responsible for separate sections of the DMFT loop, as shown in Figure~\ref{fig:ONETEP_and_TOSCAM}. As the calculation proceeds, these two programs are called in alternation, with the entire procedure being driven by an overarching script.

This splitting makes our algorithm highly amenable to parallelization: parallel TOSCAM instances can consider different correlated subspaces in isolation. That is, a system with many correlated sites is embarrassingly parallel if inter-site correlation can be neglected. By design, the AIM solver in TOSCAM is as modular as possible. This allows for it to be easily interchanged with other solvers that have been independently developed.

TOSCAM is freely available; email \texttt{cedric.weber@kcl.ac.uk} to be given access to the git repository. Note that it is dependent on ONETEP (version 5.0 and later), which can be obtained separately (see \url{www.onetep.com}).

\subsection{Scaling}
\label{sec:toscam_scaling}
\begin{figure}[!t]
  \centering
  \includegraphics[width=\linewidth]{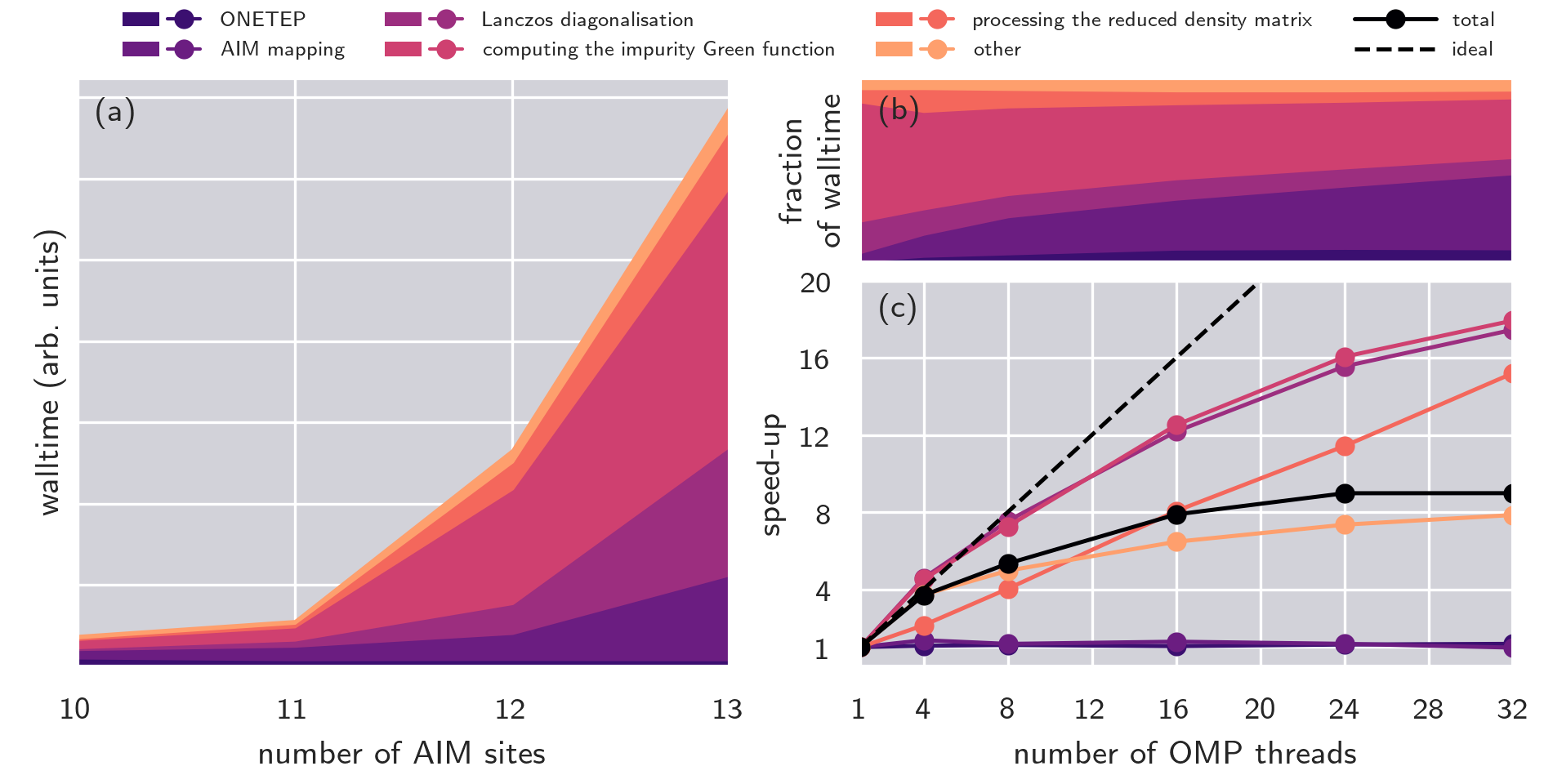}
  \caption[The scaling of {ONETEP}+{TOSCAM} with respect to the number of {AIM} sites and the number of {OpenMP} threads]
  {The scaling of {ONETEP}+{TOSCAM} for calculations on iron porphyrin (see Section~\ref{sec:application} for details). 
  (a) The scaling with respect to the number of AIM sites; (b) and (c) the fractional wall time and the speed-up with respect to the number of OpenMP threads. 
  ``Lanczos diagonalisation'' and ``computing the impurity Green's function'' are two steps involved in solving the AIM; for details refer to the Supplementary Information.}
  \label{fig:toscam_scaling}
\end{figure}

One of our primary considerations is how ONETEP+TOSCAM calculations scale. As discussed already, obtaining the Green's function of the AIM scales very poorly with the number of AIM sites. This is shown in Fig.~\ref{fig:toscam_scaling}a. We are not entirely in a position to dictate the number of AIM sites: a $3d$ correlated site is represented as a five-site impurity, and typically we need to include at least six bath sites to give the AIM sufficient flexibility to fit the hybridisation function.

To some extent, poor scaling can be overcome by efficient parallelization. Both ONETEP and TOSCAM employ hybrid MPI and OpenMP parallelization schemes. ONETEP's parallelization is highly optimized. Individual atoms are distributed across MPI threads, with lower-level computationally-intensive operations (including 3D FFT box operations, sparse matrix algebra operations, calculation of integrals, and Ewald summation) being further parallelized with OpenMP.\cite{Wilkinson2014}

In the implementation of TOSCAM, individual MPI tasks are responsible for individual correlated atoms. For systems where we have only one unique correlated atom, MPI becomes redundant. Meanwhile, OpenMP is deployed to speed up lower-level operations (see Fig.~\ref{fig:toscam_scaling}b and c).

\section{Iron porphyrin}
\label{sec:application}

\begin{figure}[!t]
  \centering
  \begin{subfigure}[!t]{0.4\columnwidth}
    \includegraphics[height=\columnwidth, angle=96, origin=c]{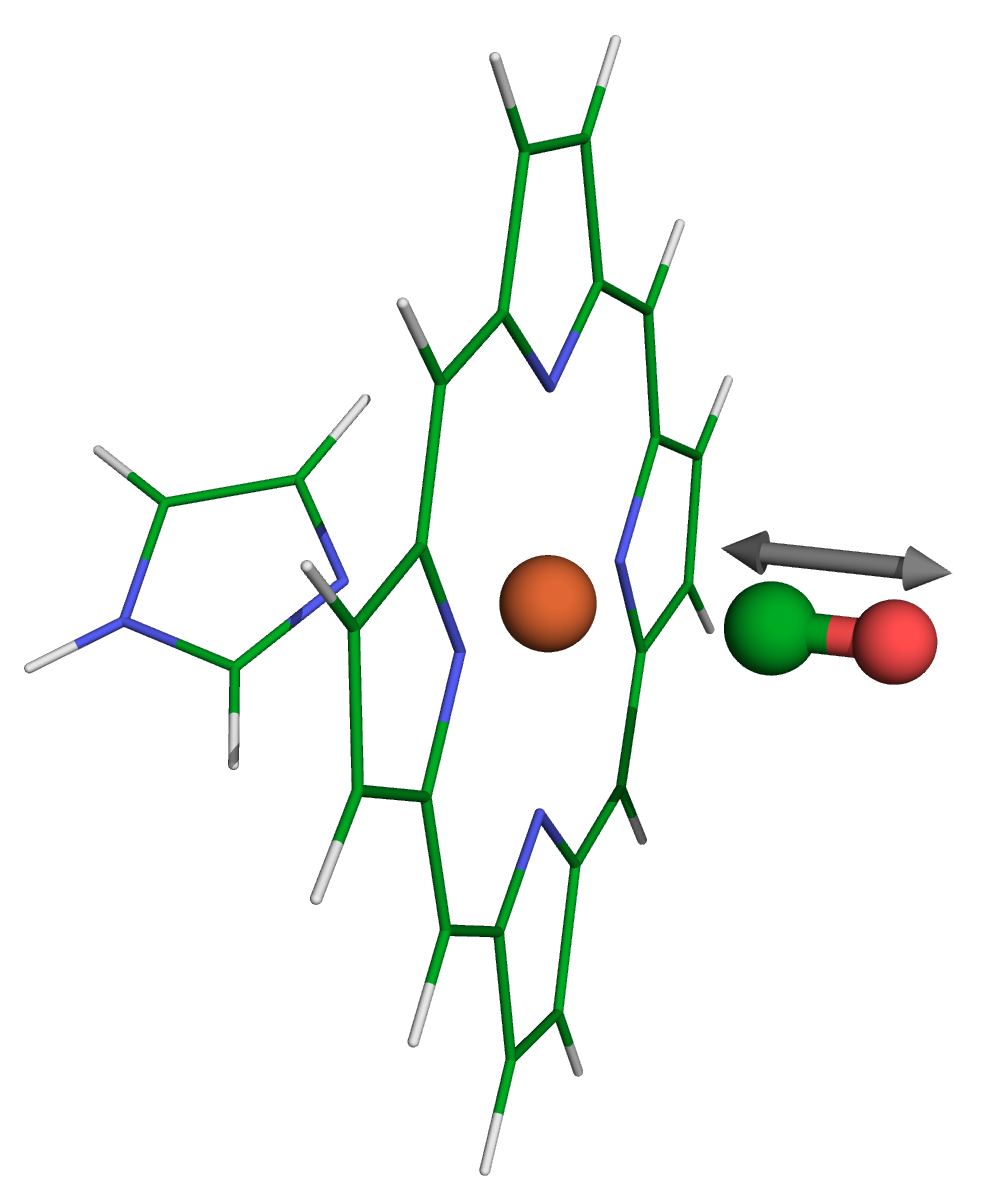}
    \caption{}
  \end{subfigure}
  \hspace{0.05\columnwidth}
  \begin{subfigure}[!t]{0.5\columnwidth}
    \centering
    
\newlength{\myimscale}
\setlength{\myimscale}{0.04\textwidth}
\begin{tikzpicture}
 \node[inner sep=0pt] (myoglobin) at (0\myimscale,0\myimscale)
     {\includegraphics[width=12\myimscale]{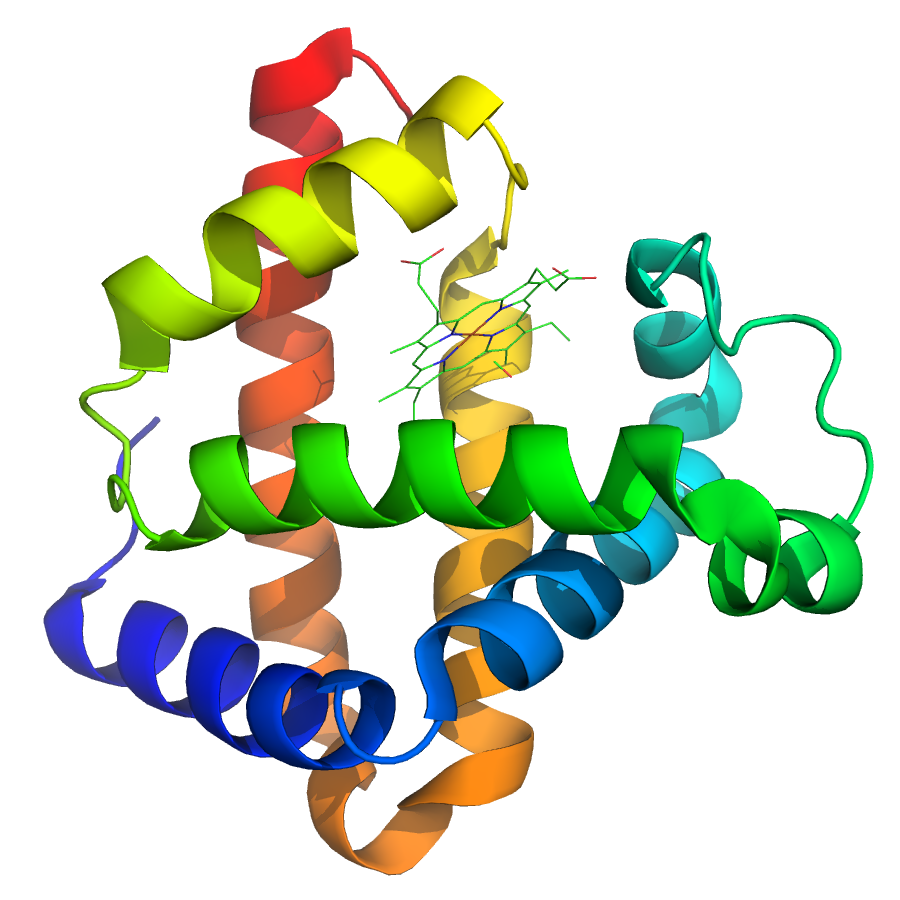}};
 \node[circle, draw, thick, minimum size=14\myimscale, inner sep=0, outer sep=0,
            path picture={
                \node at (path picture bounding box.center){
                    \includegraphics[width=15\myimscale]{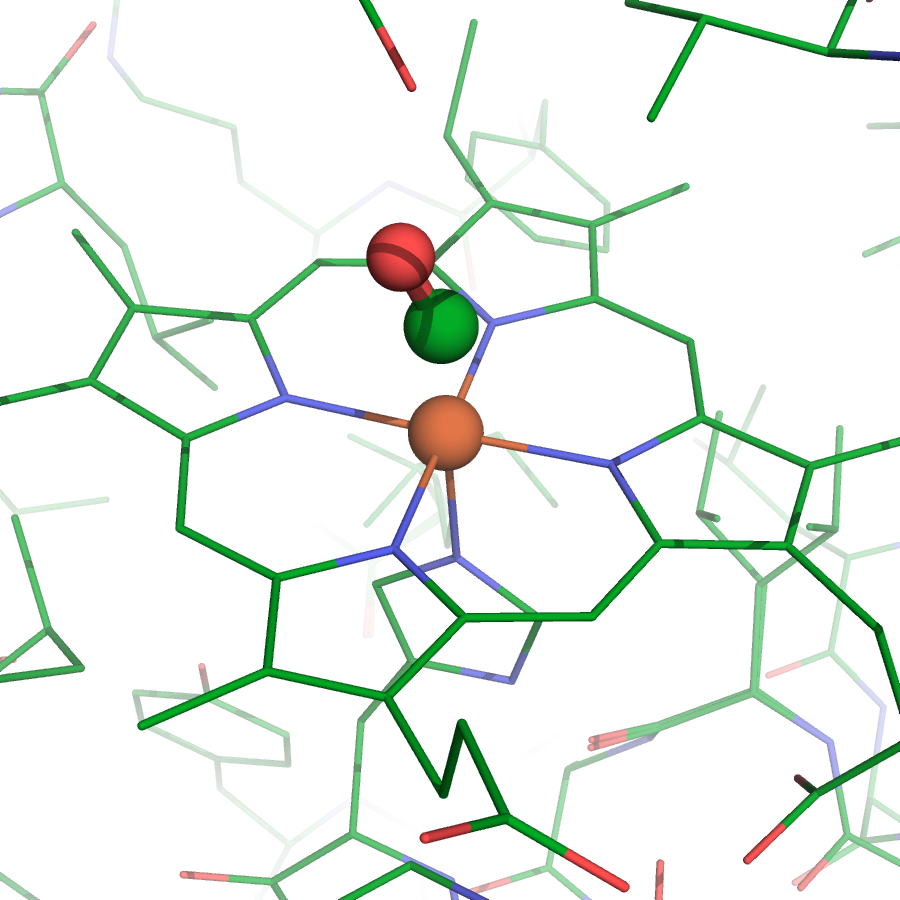}
                };
            }] (big_m) at (12\myimscale,-5\myimscale) {};
 \node[draw,circle, thick, minimum size=3\myimscale, inner sep=0, outer sep=0] (small_m) at (0.4\myimscale,1.5\myimscale) {};
 \draw[thick] (tangent cs:node=small_m,point={(big_m.north)},solution=1) -- (tangent cs:node=big_m,point={(small_m.north)},solution=2);
 \draw[thick] (tangent cs:node=small_m,point={(big_m.west)},solution=2) -- (tangent cs:node=big_m,point={(small_m.west)},solution=1);
\end{tikzpicture}
    \caption{}
  \end{subfigure}
  \caption[Carboxymyoglobin and {FePImCO}]{(a) The model complex studied in this work: iron porphyrin with axial imidazole and carbon monoxide ligands. Hydrogen, carbon, nitrogen, oxygen and iron atoms are shown in white, green, blue, red, and orange respectively. (b) Carboxymyoglobin, showing the iron binding site.\cite{Bourgeois2003} }
  \label{fig:heme}
\end{figure}
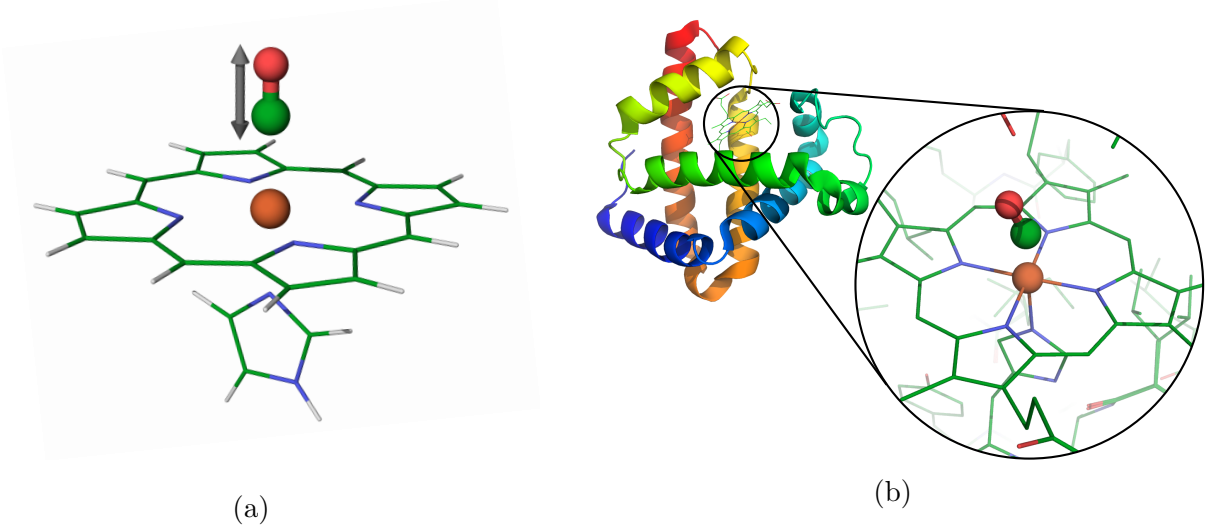

To demonstrate the use of the ONETEP+TOSCAM interface, the second half of this paper presents some calculations on an archetypal strongly-correlated system: an iron porphyrin ring with imidazole and carbon monoxide as the axial ligands (FePImCO) shown in Fig.~\ref{fig:heme}a, a toy model for the full carboxymyoglobin complex (Fig.~\ref{fig:heme}b). By translating the carbon monoxide molecule perpendicular to the porphyrin plane, we model the photodissociation of carboxymyoglobin. Myoglobin is one of the most ubiquitous metalloproteins. Previous studies have successfully applied DMFT in order to rationalise its binding energetics,\cite{Weber2013,Weber2014a} and there are unresolved questions surrounding the process of carbon monoxide photodissociation, as we shall discuss.

\subsection{Computational details}\label{subsec:dmft_heme_computational_details}
All DFT calculations were performed using a modified copy of ONETEP.\cite{Skylaris2005a, Hine2011a, ORegan2012a, ORegan2010a, Ruiz-Serrano2012a, ORegan2011a} Those modifications were subsequently included in ONETEP 5.0. All calculations used the PBE XC functional,\cite{Perdew1996a} were spin-unpolarised, and had an energy cut-off of 908\,eV. There were thirteen NGWFs on the iron atom, four on each carbon, nitrogen, and oxygen, and one on each hydrogen. All NGWFs had 6.6\,\AA\ cut-off radii. Open boundary conditions were achieved using a padded cell and a spherical Coulomb cut-off.\cite{Hine2011b} Scalar relativistic pseudopotentials were used, generated in-house using OPIUM,\cite{opium,Kerker1980a,Kleinman1982a,Hamann1989a,Rappe1990a,Gonze1991a,Ramer1999a,Grinberg2000a} and the Hubbard projectors were constructed from the Kohn-Sham solutions for a lone iron pseudopotential.\cite{Ruiz-Serrano2012a}

The bound structure was taken from Ref.~\onlinecite{Kochman2016a}, which had been optimized with the B3LYP functional. The other structures were generated by simply translating the carbon monoxide molecule in steps of 0.1\AA, without subsequently performing a geometry optimization of the rest of the system. (An ideal analysis would involve a constrained geometry optimization, to account for effects such as doming.)

Both charge-conserving and self-consistent calculations were performed, using enlarged AIM Hamiltonians via the cluster perturbation theory (CPT) extension. \edit{Seven} bath orbitals proved necessary for the AIM to be able to fit the hybridization function using the BFGS minimisation algorithm, and the AIM was solved using an ED Lanczos solver. Values of $U = 4.0$\,eV and $J=0.7$\,eV were used in the AIM Hamiltonian. \edit{The DMFT calculations were deemed to have converged when (a) the chemical potential changed by less than 1 mHa, (b) the total number of electrons was within 0.01e of the target value, and (c) the occupancy of the correlated subspace changed by less than 0.01 electrons from one iteration to the next.}

Example input and output files can be found on Materials Cloud.\cite{Linscott2020b}

\subsection{The quantum-mechanical state of the \emph{3d} iron subspace}
\label{sec:dmft_heme_reduced_density_matrix}
\begin{figure}[!t]
  \centering
  \includegraphics[width=4in]{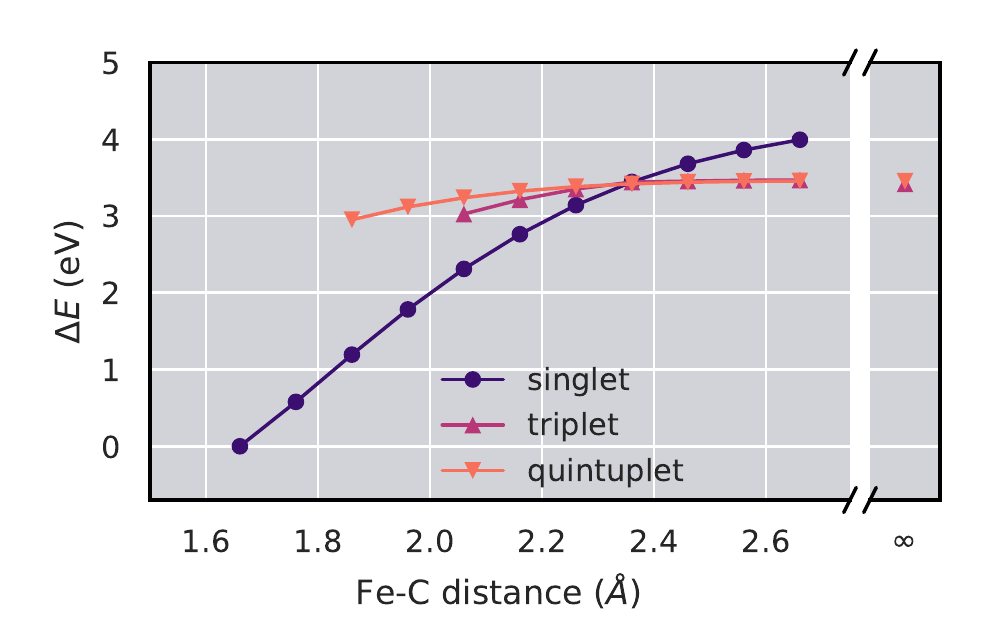}
  \caption[DFT spin state energetics of {FePImCO} during CO dissociation]{Spin state energies as given by {DFT}. For {FePImCO} the singlet state is correctly preferred, but for {FePIm} the triplet is wrongly preferred, albeit only very marginally (by 34\,meV).}
  \label{fig:heme_DFT_spin_states}
\end{figure}

A large effort (largely in the quantum chemistry community) has been made to correctly predict the spin state of Fe(II)P with (and without) a variety of axial ligands. These range from decades-old Hartree-Fock calculations to recent FCIQMC studies.\cite{Obara1982,Choe1998,Choe1999,Pierloot2003,Groenhof2005,Manni2016} FePImCO is one of the simpler cases, with a singlet state universally predicted. Meanwhile, FePIm has proven to be more of a challenge. Experiment characterises {FePIm} as a quintet. Semi-local DFT wrongly predicts it to be a triplet (as shown in Fig.~\ref{fig:heme_DFT_spin_states}). DFT\,+\,$U$ remedies this,\cite{Scherlis2007a} as does Hartree-Fock (HF).\cite{Obara1982}

\begin{figure}[!t]
  \centering
  \includegraphics[width=\textwidth]{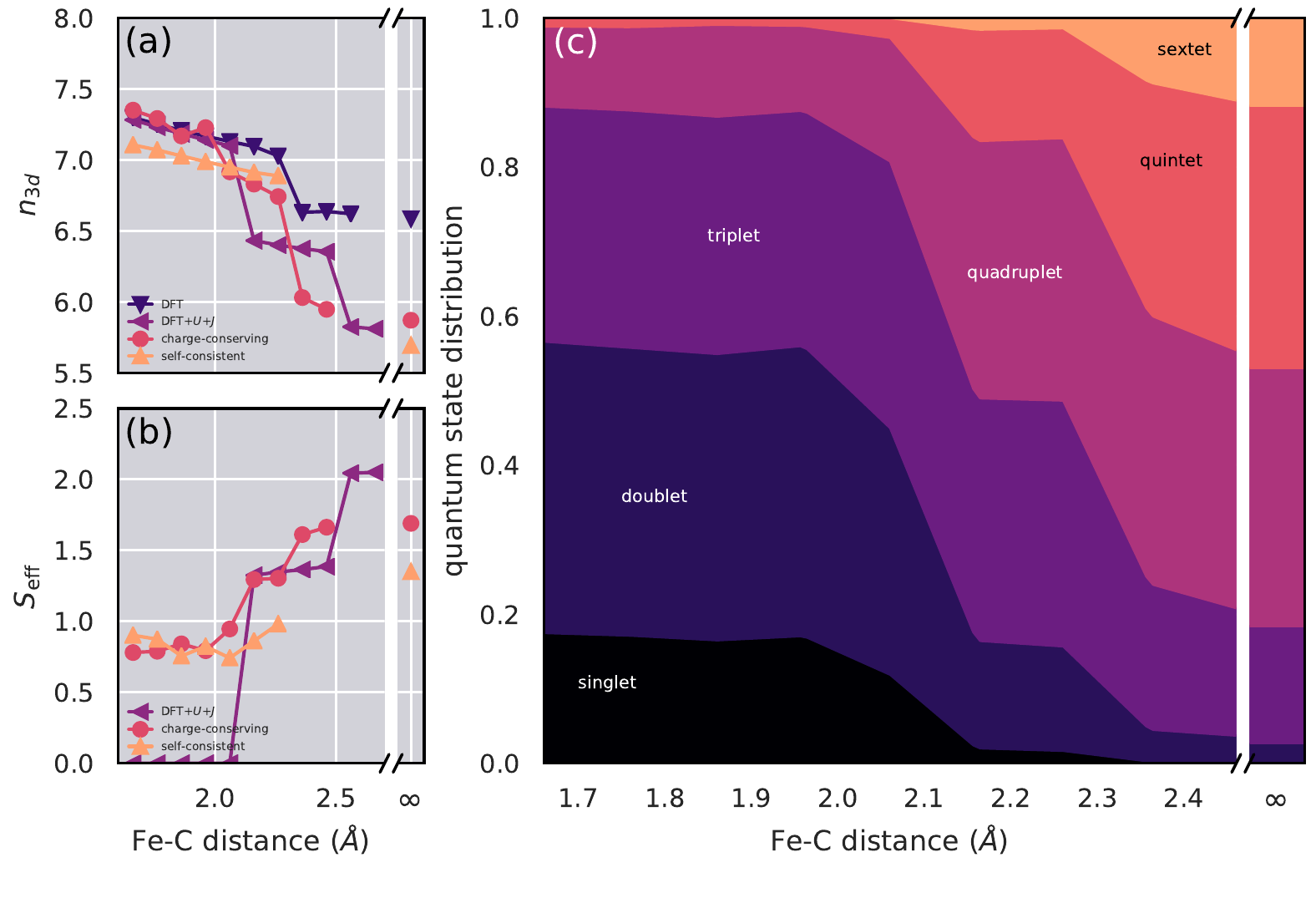}
  \caption[The electronic state of iron in {FePImCO} during {CO} dissociation]{The electronic state of iron in {FePImCO} during CO dissociation. (a) The total occupancy of the Fe-$3d$ subspace as given by {DFT} and two different {DMFT} schemes. Unfortunately self-consistent {DMFT} calculations proved very difficult to converge beyond the low-to-high spin transition, so these results have been excluded throughout. Below this transition, the two methods qualitatively agree. (b) The effective spin $S_\mathrm{eff}$ of the reduced density matrix, defined via $\mathrm{Tr}[\hat S^2 \hat \rho] = \hbar^2 S_\mathrm{eff}(S_\mathrm{eff} + 1$) \edit{for the DMFT calculations and as $S_\mathrm{eff} = \frac{1}{2}(n^\uparrow_{3d} - n^\downarrow_{3d})$ for the broken-symmetry DFT+$U$+$J$ calculations}. (c) The decomposition of the reduced density matrix by spin state. The colours correspond to the respective weights of the different contributions; if a colour occupied all the vertical axis, it would mean that all eigenvectors of the density matrix are in that particular quantum sector.}
  \label{fig:heme_reduced_density_matrix_analysis}
\end{figure}

To start, we will examine the charge transfer that takes place during CO dissociation in the DFT\,+\,DMFT picture. The Fe atom in {FePIm} is formally in the 2+ state ($d^6$). When it binds CO, it moves closer to 1+ ($d^7$) due to ligand-to-metal charge transfer. This is corroborated by our {DFT}+{DMFT} calculations: the occupancy of the $3d$ subspace can be calculated via

\begin{equation}
n_{3d} = \frac{1}{2\pi i} \sum_m \int \mathrm{d}\omega \, {G_\mathrm{imp}}_{mm}(\omega) - {{G_\mathrm{imp}}_{mm}}^\dag(\omega).
\label{eqn:subspace_occupancy_via_Gimp}
\end{equation}
This is plotted in Fig.~\ref{fig:heme_reduced_density_matrix_analysis}a as a function of the Fe--C distance. The unbinding is plainly visible as a sudden step in the total occupancy, at the same distance that {DFT} predicted the low-to-high-spin crossover (refer back to Fig.~\ref{fig:heme_DFT_spin_states}). The effect of {DMFT} is especially pronounced at large Fe-C distances, where it drives the subspace occupancy towards the expected formal $d^6$ configuration. (In some sense, {DMFT} restores the quantized nature of the electrons in the correlated subspace that is absent in DFT.)

As a means of analysing the spin state of the iron atom during the dissociation process with {DMFT}, we construct the reduced density matrix
\begin{equation}
\hat \rho = \sum_i e^{-\beta E_i} \mathrm{Tr}_B[\ket{i}\bra{i}],
\end{equation}
where we take the partial trace of the low-lying eigenstates of the {AIM} over the bath degrees of freedom, leaving a mixed density operator for the impurity alone. It is then straightforward to calculate the expectation value of $\hat S^2 = \sum_{i,j} \hat{\mathbf{S}}_i \cdot \hat{\mathbf{S}}_j$ and extract the effective spin $S_\mathrm{eff}$ (Fig.~\ref{fig:heme_reduced_density_matrix_analysis}b). Here we can see that at large distances we approach the quintet $S_\mathrm{eff} = 2$. At small distances we are closer to the triplet value $S_\mathrm{eff} = 1$. Note that this does not mean that {DMFT} has failed to predict that {FePImCO} is a singlet. Rather, this result is compatible with (but does not confirm the existence of) a singlet forming across the Fe-CO bond. By limiting ourselves to the Fe subspace we cannot directly measure such a singlet.

\edit{For comparison, we also present the occupancy and spin of the $3d$ subspace as calculated by DFT+$U$+$J$, using the same $U$ and $J$ parameters as for the DFT+DMFT calculations. DFT+$U$(+$J$) is a widespread and computationally cheap correction to DFT for accounting for electronic correlation, and is equivalent to solving the DMFT impurity problem at the level of Hartree-Fock.\cite{Anisimov1991a,Anisimov1997a,Himmetoglu2011a,ORegan2010a} We note that DFT+$U$+$J$ recovers the correct quintet spin state in the dissociated limit, in line with previous DFT+$U$ studies\cite{Scherlis2007a}, with a window where the triplet state is favored. Furthermore, the singlet state becomes unstable at a shorter Fe--C distance of around 2~\AA{} (see Fig.~\ref{fig:heme_reduced_density_matrix_analysis}b). This is a common feature of DFT+$U$,\cite{Kulik2011a,Linscott2018} where the corrective $+U$ potential reduces the hybridization between the correlated subspace and the ligand orbitals, thereby weakening the bond between them. Addressing this within a Hubbard model framework requires more sophisticated approaches such as inter-site terms\cite{Kulik2011a} or applying corrective Hubbard potentials to ligand subspaces.\cite{Linscott2018} This is not a problem that DMFT suffers from.}

To inspect the \edit{DMFT} reduced density matrix in more detail, one can construct the spin-projector
\begin{equation}
\hat P_S = \sum_ {s \in S} \ket{s} \bra{s}
\end{equation}
as the sum of the eigenstates $\ket{s}$ of the operator $\hat S^2$ with eigenvalue $S(S + 1)$. This allows us to evaluate the fraction of the reduced density matrix in singlet, doublet, triplet, and higher states via $\Trace{\hat P_S \hat \rho \hat P_S}$ for $S = 0$, $\frac{1}{2}$, $1$ etc. Note, however, that this approach is incompatible with the {CPT} extension. The {CPT} extension involves solving an auxiliary {AIM} Hamiltonian that shares the same impurity Green's function as a larger {AIM} Hamiltonian, and consequently any quantities derived directly from the Green's function will be unaffected. However, there is no such guarantee for the reduced density matrix, because the hybridization function of this auxiliary system does not necessarily match that of the physical system. To overcome this, the {CPT} extension was at first applied in order to obtain an approximate solution, but then removed for the final {DMFT} step. Typically this final step required the addition of an extra bath site so that the {AIM} acquired sufficient flexibility to fit the impurity hybridization function to the local hybridization function without the assistance of the {CPT} extension.

The decomposition of the reduced density matrix into spin sectors is displayed in Fig.~\ref{fig:heme_reduced_density_matrix_analysis}c. It reveals a large quintet state contribution in the limit of dissociation, but also that, regardless of Fe-C distance, many different spin sectors are important. This would be missed if we only examined $S_\mathrm{eff}$ or only performed DFT. %

Evidently, a multitude of states play an important role throughout CO-unbinding, and therefore the success of {DFT\,+\,$U$} and {HF} in predicting the quintet ground state must be for the wrong reasons, as neither go beyond the single-determinantal picture. (Note that {HF} is known to overly favour high-spin states.\cite{Neese2009})

It should be noted that the precise details of Fig.~\ref{fig:heme_reduced_density_matrix_analysis} are somewhat sensitive to various simulation parameters --- most notably the definition of the Hubbard projectors --- but qualitatively the results are expected to hold generally.

\subsection{Photodissociation}
The photodissociation mechanism of carboxymyoglobin is already relatively well understood. Irradiation at 570\,nm (2.18\,eV) causes the excitation of electrons in the porphyrin ring into low lying singlet states with $\pi$/$\pi^*$ character (the so-called Q band).\cite{Franzen2001a} The carbon monoxide ligand then dissociates within 50\,fs, as the system adiabatically crosses to a repulsive anti-back-bonding orbital.\cite{Dreuw2002a, Dunietz2003a} There is a small (but not insignificant) predicted energy barrier of 0.08\,eV between these two states, as calculated by {B3LYP} and {TDDFT}.\cite{Kochman2016a} The porphyrin then undergoes the ``intersystem crossing", a complicated, multi-step process which ultimately takes the dissociated system to its high-spin ground state.

\begin{figure}[t]
   \centering
   \includegraphics[width=4in]{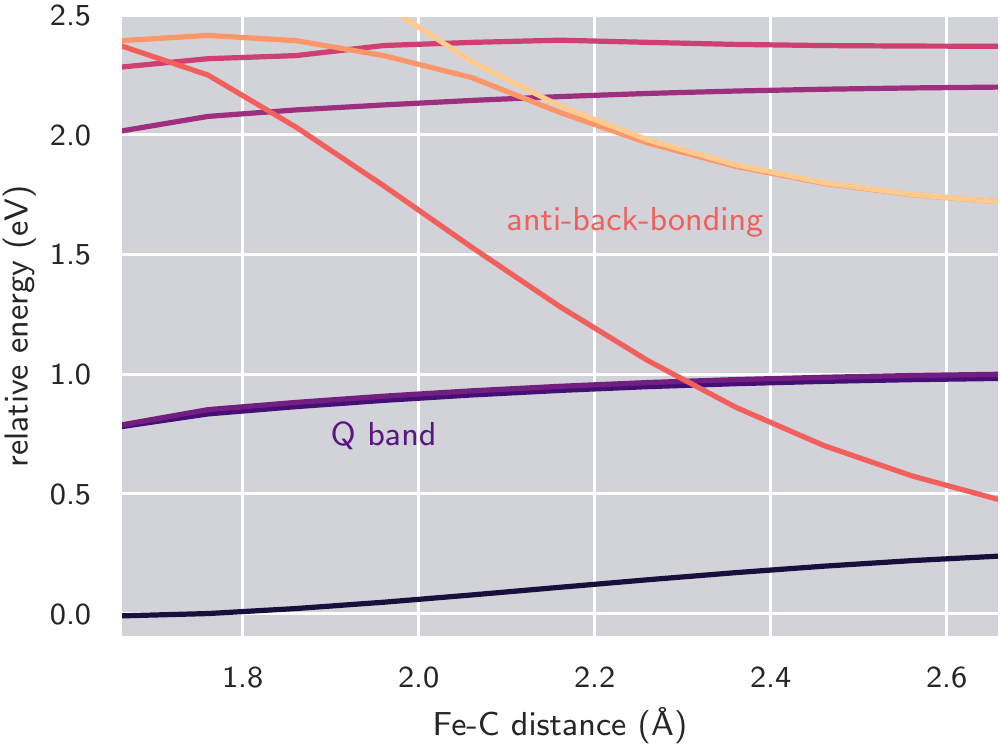}
   \caption[The Kohn-Sham eigenvalues of {FePImCO} during CO dissociation]{Energies of the {KS} molecular orbitals, measured relative to the highest occupied orbital of the tightly-CO-bound structure.}
   \label{fig:heme_bands_DFT}
\end{figure}

Semi-local {DFT} captures this process qualitatively. The energies of the lowest unoccupied {KS} molecular orbitals as predicted via {DFT} are shown in Fig.~\ref{fig:heme_bands_DFT}. The Q band is present, and the pathway from the Q band to the anti-back-bonding orbital is clearly visible via their crossing at approximately 2.3\,\AA\ (the same distance we observe the low-to-high spin crossover in Fig.~\ref{fig:heme_DFT_spin_states}), with an energy barrier of approximately 0.13\,eV. Compared to the {TDDFT}/{B3LYP} results of Refs.~\onlinecite{Dreuw2002a} and \onlinecite{Dunietz2003a}, {PBE} calculations place this crossover at a much longer distance (approximately 2.3\,\AA\ compared to 2.0\,\AA), and predict that the energy of the anti-back-bonding orbital drops much more steeply. (Head-Gordon and co-workers noted that the very gentle decrease in the energy of the anti-back-bonding orbital as predicted by their {TDDFT}/{B3LYP} calculations is at odds with the $\sim50$\,fs timescale of photodissociation.\cite{Dunietz2003a})

\begin{figure}[!t]
  \centering
  \includegraphics[width=3in]{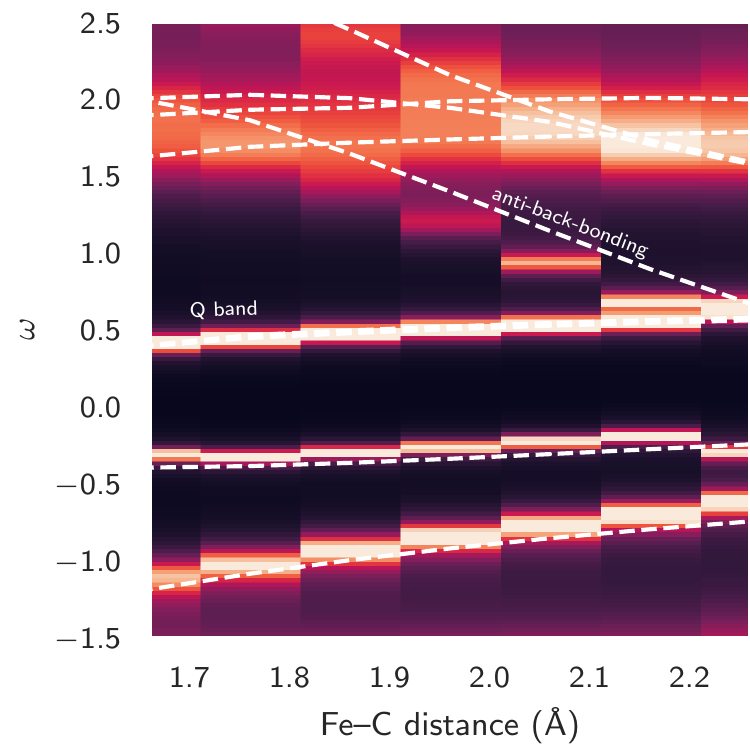}
  \caption[The {DMFT} {DOS} of {FePImCO} during dissociation, compared to the {KS} eigenenergies]{The {DMFT} {DOS} of {FePImCO} during dissociation, compared to the {KS} eigenenergies (white dashed lines), as given by self-consistent {DMFT} calculations. The DOS and eigenenergies have been aligned to match the Q band, because, being a porphyrin-ring state, it should not be significantly shifted by {DMFT}.}
  \label{fig:heme_DMFT_bands}
\end{figure}

To compare the results of {DMFT} to these {KS} eigenenergies, the analogous quantity we must extract is the {DOS}. The {DOS} is given by the trace of the many-body density matrix
\begin{equation}
\rho(\omega) = \sum_{\alpha, \beta} \rho^{\alpha \beta}(\omega) S_{\beta \alpha}.
\end{equation}
The {DMFT} {DOS} is compared to the {KS} eigenenergies in Fig.~\ref{fig:heme_DMFT_bands}. Qualitatively, they yield very similar results, but with DMFT (unlike DFT) we obtain the lifetime of the excitations.

\begin{figure}[!t]
  \centering
  \includegraphics[width=4in]{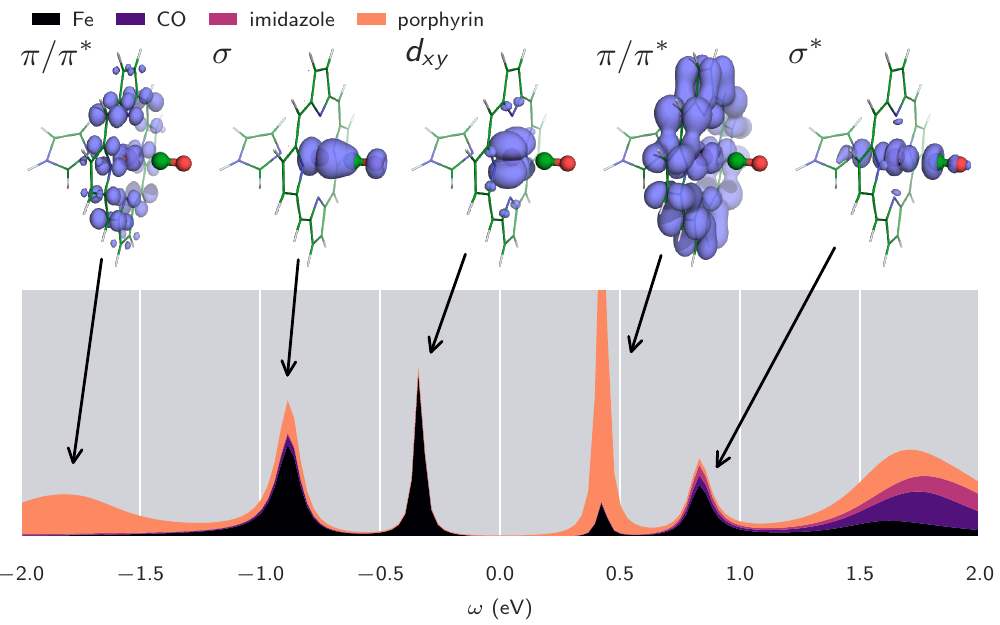}
  \caption[The {DMFT} density of states for FePImCO]{Self-consistent DMFT density of states for carboxy-heme with a Fe-C distance of 2.06\,\AA. The DOS is further decomposed into contributions from the iron atom, CO molecule, imidazole ligand and porphyrin ligand. Above, isosurfaces of $\rho(\br,\omega_\text{peak})$ have been plotted for each peak.}
  \label{fig:heme_DMFT_ldos_with_isosurfaces}
\end{figure}

To reveal the contribution of individual atoms (or groups of atoms) towards the {DMFT} {DOS}, it can be decomposed into local densities of state (LDOSs)
\begin{equation}
\rho_I(\omega) = \sum_{\alpha \in I} \sum_\beta \rho^{\alpha \beta}(\omega) S_{\beta \alpha},
\end{equation}
where $I$ denotes a subset of NGWFs typically belonging to atoms that are a particular element or part of a spatially distinct subsystem (\emph{e.g.} all the NGWFs belonging to atoms in the porphyrin ring). One such LDOS is plotted in Fig.~\ref{fig:heme_DMFT_ldos_with_isosurfaces}, along with isosurfaces of the spectral density at energies corresponding to the various peaks in the {DOS}. The Q-band $\pi/\pi^*$ orbitals and the Fe-CO back- and anti-back-bonding orbitals are all clearly identifiable.

\begin{figure}[!t]
  \centering
  \includegraphics[width=6.3in]{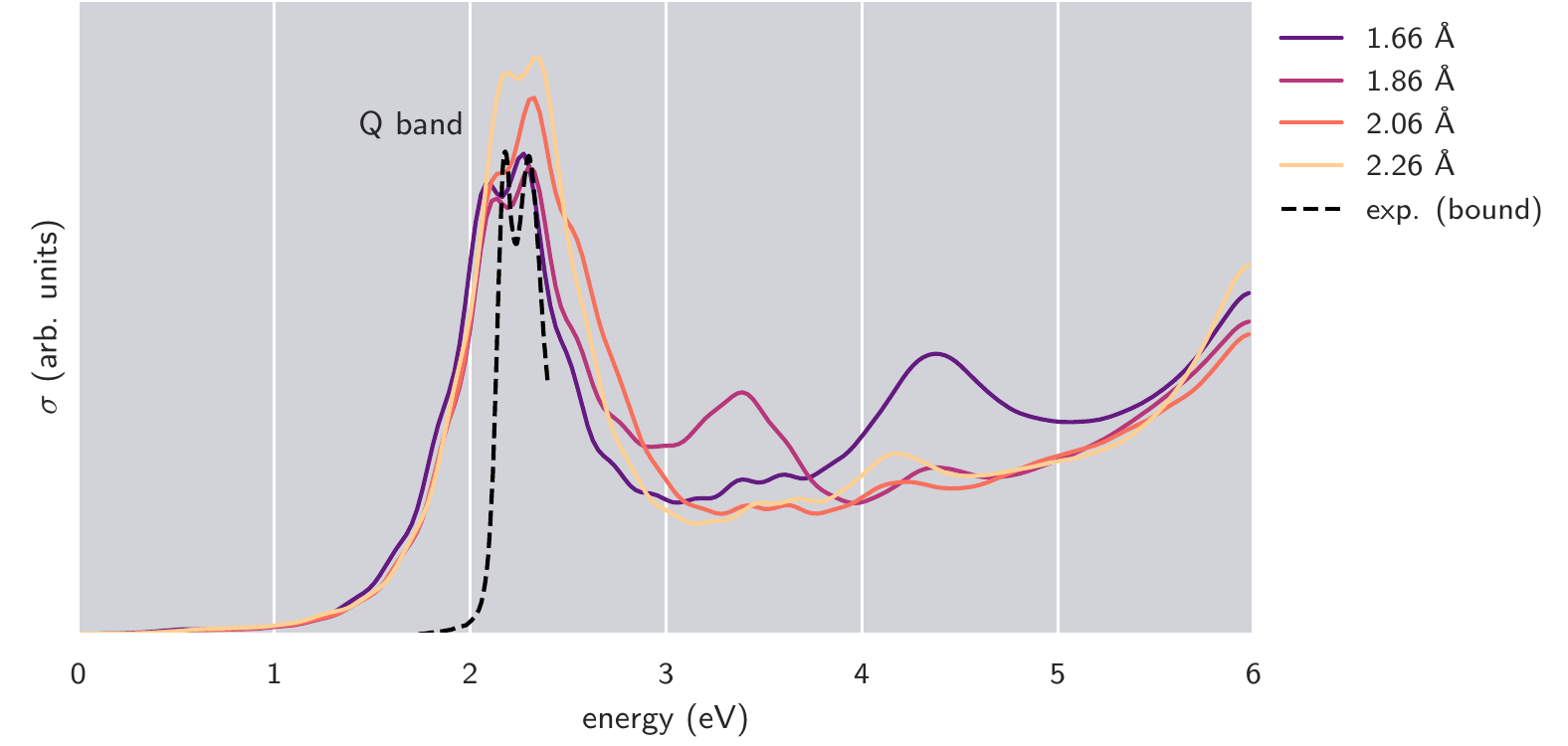}
  \caption[Optical spectra of {FePImCO}]{Optical spectra of FePImCO calculated using self-consistent DMFT, going from ligated (dark) up to the point of dissociation (light). Also pictured are the Q-band peaks from experimental spectra of carboxymyoglobin.\cite{Eaton1979}}
  \label{fig:heme_DMFT_spectra}
\end{figure}

Another important quantity that can be extracted from DMFT calculations is the optical spectrum. The theoretical optical absorption spectrum can be obtained within the linear-response regime (that is, Kubo formalism) as
\begin{equation}
\sigma_{ij}(\omega) = \frac{2\pi}{\Omega} \int d\omega' \frac{f(\omega' - \omega) - f(\omega')}{\omega}\left(\rho^{\alpha \beta}(\omega' - \omega)\mathbf{v}^i_{\beta \gamma} \rho^{\gamma \delta}(\omega') \mathbf{v}^j_{\delta \alpha}\right)
\end{equation}
where $\Omega$ the simulation cell volume, $f(\omega)$ is the Fermi-Dirac distribution, $\rho$ is the basis-resolved spectral density, the $i$ and $j$ indices correspond to Cartesian directions, the velocity operator $\mathbf{v}$ is
\begin{equation}
\mathbf{v}^j_{\alpha \beta} = -i\bra{\alpha} \nabla_j \ket{\beta} + i \bra{\alpha} \left[\hat V_{nl}, \mathbf{r}\right]\ket{\beta}
\end{equation}
which includes the effect of non-local pseudopotentials $V_{nl}$ on the velocity operator matrix elements, and adopts the no-vertex-corrections approximation.\cite{Millis2004} Optical spectra for heme are typically carried out in liquid or gas phases, and so are described by the isotropic part of the optical conductivity tensor
\begin{equation}
\sigma(\omega) = \frac{1}{3}\sum_i \sigma_{ii}(\omega).
\end{equation}
The optical absorption spectra for carboxy-heme complexes as given by self-consistent DMFT are plotted in Fig.~\ref{fig:heme_DMFT_spectra}. These spectra are dominated by a feature at around 2\,eV associated with $\pi$-$\pi^*$ transitions on the porphyrin ring --- that is, the Q band. The double-peak structure of the Q band is successfully reproduced. (Ref.~\onlinecite{Weber2014a} found that $J > 0$ is necessary to obtain this double-peak feature.) Secondary peaks appear above 3\,eV corresponding to direct photoexcitation of the anti-back-bonding orbital.

\section{Conclusions}
This paper has described how DMFT has been interfaced with linear-scaling DFT in the ONETEP+TOSCAM implementation. Crucially, for the purposes of simulating metalloproteins, this DFT\,+\,DMFT implementation does not compromise our ability to model thousands of atoms at the DFT level, opening up a new frontier for accurate simulation of complex and heterogeneous systems containing transition metals and lanthanides.

The ONETEP+TOSCAM interface will continue to be developed. In particular, \edit{work is underway to compute forces at the DFT+DMFT level (as has been achieved elsewhere\cite{Pascut2016,Plekhanov2018}),} a GPU implementation of the ED solver will be incorporated, as well as a CTQMC solver (which will allow us to solve substantially larger AIMs.) Note that it is straightforward to add additional solvers due to the modularity of the code.


Calculations on the photodissociation of carboxymyoglobin showcased the kinds of results one can extract from such a DFT\,+\,DMFT calculation on a metalloprotein, including some --- such as the mixed quantum state of the iron $3d$ subspace and the finite lifetime of excitations --- that are inaccessible via DFT. And while the calculations do not reveal any previously unknown physics, there is scope here to resolve some unanswered questions surrounding the photodissociation process. In particular, the remarkably fast rate of photodissociation ($\sim 50$\,fs) is at odds with the gentle slope of the potential energy surface (discussed above) and the predicted barrier on the order of 0.1\,eV (compared to the 0.028\,eV zero-point energy of the Fe-C stretching mode).\cite{Kochman2016a} Further study could investigate this apparent contradiction.

\acknowledgement
EBL acknowledges financial support from the Rutherford Foundation Trust and the EPSRC Centre for Doctoral Training in Computational Methods for Materials Science under grant EP/L015552/1. NDMH acknowledges the support of the EPSRC under grant EP/P02209X/1. MCP acknowledges the support of the EPSRC under grant EP/P034616/1. CW acknowledges the support of the EPSRC under grant EP/R02992X/1. The authors thank M.\ A.\ Al-Badri and M.\ J.\ Rutter for useful discussions. This work was performed using the Darwin Supercomputer of the University of Cambridge High Performance Computing Service (\url{http://www.hpc.cam.ac.uk/}), provided by Dell Inc.\ using Strategic Research Infrastructure Funding from the Higher Education Funding Council for England and funding from the Science and Technology Facilities Council.

\begin{suppinfo}
\edit{The Supporting Information contains a} brief description of the TOSCAM/ONETEP interface, as well as relevant links\edit{, and a}n introduction to the Lanczos algorithm and how it may be used to solve the Anderson impurity model.
\end{suppinfo}

\bibliography{paper_onetep_DMFT.bib}

\clearpage
\begin{center}
\textbf{\huge \sffamily Supplementary Material}
\end{center}
\setcounter{equation}{0}
\setcounter{figure}{0}
\setcounter{table}{0}
\setcounter{page}{1}
\setcounter{section}{0}
\makeatletter
\renewcommand{\theequation}{S\arabic{equation}}
\renewcommand{\thefigure}{S\arabic{figure}}
\renewcommand{\bibnumfmt}[1]{[S#1]}
\renewcommand{\citenumfont}[1]{S#1}

\section{Accessing the codes}

ONETEP is available under academic license to all UK academics and is also part of Materials Studio (\url{http://accelrys.com/products/datasheets/onetep.pdf}). For more details see \url{www.onetep.com}.

To obtain the source code for TOSCAM, contact C\'edric Weber to be granted access to the git repository. TOSCAM is distributed under the lesser GNU public license.

\section{Exact Diagonalization}
\subsection{The standard Lanczos algorithm}
The Lanczos method is an approach for obtaining the eigenvectors and eigen-energies of a Hermitian matrix $A$, without ever having to perform a full diagonalisation. Starting with some arbitrary normalised vector $\ket{0}$, we compute $\epsilon_0 = \bra{0}A \ket{0}$. Then we construct $\tilde{\ket{1}} = \hat A\ket{0}-\epsilon_0 \ket{0}$, and normalise to obtain $\ket{1}$. Importantly, the resulting vector $\ket{1}$ is orthogonal to $\ket{0}$.

We can now generate a third vector $\tilde{\ket{2}} = A\ket{1}-\epsilon_1\ket{1} - k_1\ket{0}$, where $k_1 = \bra{0}A \ket{1}$, and normalise to obtain $\ket{2}$. Again, $\ket{0}$, $\ket{1}$, and $\ket{2}$ are orthogonal by construction.

Now suppose we were to continue to generate orthogonal vectors according to this pattern 
\begin{equation}
\ket{i+1} = \frac{1}{\sqrt{\bra{i}(A - \epsilon_i)^2\ket{i} + {k_i}^2}} A \ket{i} - \epsilon_i\ket{i} -k_i \ket{i-1}
\end{equation}
to obtain a basis of Lanczos vectors $\{\ket{i}\}$. In this basis, the matrix $A$ is tridiagonal:\footnote{This is straightforward to show. For example, $ \bra{j}A\ket{i} = \bra{j}\left( \tilde{\ket{i+1}}+\epsilon_i\ket{i}+k_i\ket{i-1}\rangle \right) = 0$ if $i \leq j-2 $. The other entries can be obtained via similar logic.}
\begin{equation}
A_{ij} = \begin{pmatrix}
\epsilon_0 & k_1        & 0          & \cdots     & \cdots \\
k_1        & \epsilon_1 & k_2        & 0          & \cdots \\
0          & k_2        & \epsilon_2 & k_3        & 0      \\
\vdots     & 0          & k_3        & \epsilon_3 & \ddots \\
\vdots     & \vdots     & 0          & \ddots     & \ddots
\end{pmatrix}_{ij}.
\end{equation}
From here, it is straightforward to calculate the eigenvectors and eigenvalues of $A$.

As an approximate scheme, one need only consider the first $L+1$ Lanczos vectors. In this case, $\tilde A_{ij} = \sum^L_{kl}\braket{i}{k}\bra{k}A\ket{l}\braket{l}{j}$ is an $(L+1)$-by-$(L+1)$ tridiagonal matrix, the eigenvalue problem $\tilde Ac^\nu = E_\nu c^\nu$ is straightforward to solve, and the eigenvectors of $\tilde A$ are approximated by $\ket{\nu} = \sum_i^L c^\nu_i \ket{i}$. By progressively increasing $L$ and periodically recalculating $\{E_0,...,E_L\}$ one can converge to the eigenvectors and energies of $A$ without ever doing the full diagonalisation.

Note that this algorithm is very cheap; multiplication by $\tilde A$ is the most expensive step, and scales as $\mathcal{O}(L^2)$. It also is worthwhile noting that because the Lanczos basis is generated via repeated action of $A$ on the previous Lanczos vector, the Lanczos algorithm rapidly finds the vectors $\ket{i}$ for which $A\ket{i}$ is large --- another advantage of the method.

\subsection{Applying the Lanczos method to the AIM}
Let us now adapt the Lanczos method for the specific case of calculating the Green's function of an AIM. To calculate the diagonal components $G_\text{imp}^{\alpha\alpha}(\omega)$ we encounter terms of the form
\begin{align*}
G_\text{imp}^{\alpha\beta}(\omega) = \left\langle \psi_0 \left| \hat c_\alpha \frac{1}{\omega^+-(\hat H - E_0)} \hat c^\dag_\beta \right| \psi_0 \right\rangle
  + \left\langle \psi_0 \left| \hat c^\dag_\beta \frac{1}{\omega^++(\hat H - E_0)} \hat c_\alpha \right| \psi_0 \right\rangle.
\end{align*}
Obtaining $\ket{\psi_0}$ is straightforward: we can obtain it by performing the Lanczos algorithm on $\hat H$, as described in the previous section. Given $\ket{\psi_0}$, some additional tricks are necessary to arrive at the Green's function. Let us first focus on the diagonal components $G_\text{imp}^{\alpha\alpha}[\omega]$, in which case we are interested in quantities of the form
\begin{equation}
\left\langle \psi_0 \left| \mathcal{O}^\dag \frac{1}{z-H} \mathcal{O} \right| \psi_0 \right\rangle.
\label{eqn:DMFT_Lanczos_Gimp_simplified_term}
\end{equation}
for some generic operator $\mathcal{O}$. To calculate this, we perform the Lanczos algorithm on $H$ --- but now, instead of starting with a random vector, we choose
\begin{equation}
\ket{0} = \frac{\mathcal{O}\ket{\psi_0}}{\sqrt{\bra{\psi_0}\mathcal{O}^\dag \mathcal{O}\ket{\psi_0}}}.
\label{eqn:DMFT_Lanczos_initial_vector}
\end{equation}
%
In the Lanczos basis generated using this vector, we have
\begin{equation}
(z-H)_{ij} = \begin{pmatrix}
z-\epsilon_0 & -k_1         & 0            & \cdots       & \cdots \\
-k_1         & z-\epsilon_1 & -k_2         & 0            & \cdots \\
0            & -k_2         & z-\epsilon_2 & -k_3         & 0      \\
\vdots       & 0            & -k_3         & z-\epsilon_3 & \ddots \\
\vdots       & \vdots       & 0            & \ddots       & \ddots
\end{pmatrix}_{ij}
\end{equation}
Crucially, the quantity we ultimately want to obtain (equation \ref{eqn:DMFT_Lanczos_Gimp_simplified_term}) is $(z-H)^{-1}_{00}$, which is given\footnote{
The $ij$-element of the inverse of $A$ is given by
\begin{equation}
(A^{-1})_{ij} = (-1)^{i+j}\frac{\det \Delta_{ij}}{\det A}
\end{equation}
where $\Delta_{ij}$ is the sub-matrix of $A$ obtained by eliminating from $A$ the $i$-th row and $j$-th column. In the case of a tridiagonal matrix,
\begin{equation}
\tiny
\det A = \det \begin{pmatrix}
A_{00} & A_{01} & 0      &        &        \\
A_{10} & A_{11} & A_{12} & 0      &        \\
0      & A_{21} & A_{22} & A_{23} & 0      \\
       & 0      & A_{32} & A_{33} & A_{34} \\
       &        & 0      & A_{43} & A_{44}
\end{pmatrix}
= A_{00}\det
\begin{pmatrix}
A_{11} & A_{12} & 0      &        \\
A_{21} & A_{22} & A_{23} & 0      \\
0      & A_{32} & A_{33} & A_{34} \\
       & 0      & A_{43} & A_{44}
\end{pmatrix}
-A_{01}A_{10}\det
\begin{pmatrix}
A_{22} & A_{23} & 0      \\
A_{32} & A_{33} & A_{34} \\
0      & A_{43} & A_{44}
\end{pmatrix}.
\footnotesize
\end{equation}
If $D_i$ is determinant of the matrix $A$ having removed the first $i$ rows and columns, it follows that
\begin{equation}
\frac{D_0}{D_1} = \frac{A_{00}D_1 - |A_{01}|^2D_2}{D_1} = A_{00} - \frac{A_{01}A_{10}}{D_1/D_2}.
\end{equation}
This reasoning can be extended to
\begin{equation}
\frac{D_l}{D_{l+1}} = A_{ll} - \frac{|A_{ll+1}|^2}{D_{l+1}/D_{l+2}}
\end{equation}
and thus the first element of the inverse of $A$ is given by the continued fraction
\begin{equation}
(A^{-1})_{00} = \frac{1}{D_0/D_1} = \frac{1}{A_{00}-\frac{|A_{01}|^2}{A_{11}-\frac{|A_{12}|^2}{A_{22} - \cdots}}}
\end{equation}
} by the continued fraction
\begin{equation}
\frac{1}{z-\epsilon_0-\frac{|k_1|^2}{z-\epsilon_1-\frac{|k_2|^2}{z-\epsilon_2 - \cdots}}}
\end{equation}
which can be numerically evaluated (via, for example, the modified Lentz method \cite{Press1992a}). Thus we can calculate the diagonal terms $G_\text{imp}^{\alpha\alpha}[\omega]$ by setting $\mathcal{O} = \hat c_\alpha$. The off-diagonal terms, meanwhile, require some clever trickery: it can be shown \cite{Dolfen2006a} that
\begin{equation}
G^{\alpha\beta}_{imp} = \mathcal{G}^{\alpha\beta} - \frac{1}{2}\left(G^{\alpha\alpha}_{imp} + G^{\beta\beta}_{imp}\right)
\end{equation}
where $\mathcal{G}^{\alpha\beta}$ is the result of repeating the above process for the diagonal elements, but now using the initial Lanczos matrix $\mathcal{O} = \frac{1}{\sqrt{2}}\left(\hat c_\alpha + c_\beta\right)$. This avoids a vanishing denominator $\bra{\psi_0}c^\dag_\alpha c_\beta \ket{\psi_0}$ if we were to blindly proceed with the same procedure as for the diagonal elements.

\providecommand{\latin}[1]{#1}
\makeatletter
\providecommand{\doi}
  {\begingroup\let\do\@makeother\dospecials
  \catcode`\{=1 \catcode`\}=2 \doi@aux}
\providecommand{\doi@aux}[1]{\endgroup\texttt{#1}}
\makeatother
\providecommand*\mcitethebibliography{\thebibliography}
\csname @ifundefined\endcsname{endmcitethebibliography}
  {\let\endmcitethebibliography\endthebibliography}{}

\end{document}